\documentclass[]{aastex63}

\newenvironment{myfont}{\fontfamily{lmr}\selectfont}{\par}
\DeclareTextFontCommand{\textmyfont}{\myfont}

\usepackage{amsmath}
\usepackage{lineno}
\usepackage{bm}

\received{2021 June 18}
\revised{2021 June 18}
\accepted{\today}

\setwatermarkfontsize{5cm}

\shorttitle{KOI-126}
\shortauthors{Yenawine et al.}

\begin{document}
\newcommand{\ik}{{\it Kepler}}
\newcommand{\KIC}{{{\rm{KIC}}~5897826}}
\newcommand{\etal}{{\it et~al.~}}
\newcommand{\ie}{{\it i.e.~}}
\newcommand{\eg}{{\it e.g.~}}
\newcommand{\vs}{{\it vs.~}}
\newcommand{\kms}{\mbox{$km \ s^{-1}$}}
\newcommand{\Msun}{\mbox{M$_{\sun}$}}
\newcommand{\Mjup}{\mbox{M$_{Jup}$}}
\newcommand{\Mearth}{\mbox{M$_{\earth}$}}
\newcommand{\Rsun}{\mbox{R$_{\sun}$}}
\newcommand{\Rjup}{\mbox{R$_{Jup}$}}
\newcommand{\Rearth}{\mbox{R$_{\earth}$}}
\newcommand{\Lsun}{\mbox{L$_{\sun}$}}

\title{Photodynamical Modeling of the Fascinating Eclipses in the Triple-Star System KOI-126}

\correspondingauthor{Mitchell Yenawine}
\email{myenawine@sdsu.edu, yenawine@iastate.edu}

\author[0000-0001-6556-1536]{Mitchell E. Yenawine}
\affiliation{Department of Astronomy, San Diego State University, 5500 Campanile Drive, San Diego, CA 92182-1221, USA}
\affiliation{Department of Physics and Astronomy, Iowa State University, 2323 Osborn Drive, Ames, IA 50011-3160, USA}

\author[0000-0003-2381-5301]{William F. Welsh}
\affiliation{Department of Astronomy, San Diego State University, 5500 Campanile Drive, San Diego, CA 92182-1221, USA}

\author[0000-0001-9647-2886]{Jerome A. Orosz}
\affiliation{Department of Astronomy, San Diego State University, 5500 Campanile Drive, San Diego, CA 92182-1221, USA}

\author[0000-0001-6637-5401]{Allyson Bieryla}
\affiliation{Center for Astrophysics $\vert$ Harvard \& Smithsonian, 60 Garden Street, Cambridge, MA 02138, USA}

\author[0000-0001-9662-3496]{William D. Cochran}
\affiliation{ Center for Planetary Systems Habitability and McDonald Observatory, The University of Texas at Austin, Austin, TX 78712, USA}

\author[0000-0002-7714-6310]{Michael Endl}
\affiliation{ Center for Planetary Systems Habitability and McDonald Observatory, The University of Texas at Austin, Austin, TX 78712, USA}

\author[0000-0001-9911-7388]{David W. Latham}
\affiliation{Center for Astrophysics $\vert$ Harvard \& Smithsonian, 60 Garden Street, Cambridge, MA 02138, USA}

\author[0000-0002-8964-8377]{Samuel N. Quinn}
\affiliation{Center for Astrophysics $\vert$ Harvard \& Smithsonian, 60 Garden Street, Cambridge, MA 02138, USA}

\author[0000-0001-5504-9512]{Donald R. Short}
\affiliation{Department of Astronomy, San Diego State University, 5500 Campanile Drive, San Diego, CA 92182-1221, USA}

\author[0000-0002-6742-4911]{Gur Windmiller}
\affiliation{Department of Astronomy, San Diego State University, 5500 Campanile Drive, San Diego, CA 92182-1221, USA}


\begin{abstract}

We explore the fascinating eclipses and dynamics of the compact hierarchical triple star system KOI-126 (KIC 5897826).
This system is comprised of a pair of M-dwarf stars (KOI-126 B and C) in a 1.74 day orbit which revolve around an F-star (KOI-126 A) every 34 days. 
Complex eclipse shapes are created as the M stars transit the F star, due to two effects: (i) the duration of the eclipse is a significant fraction of the M-star orbital period, so the prograde or retrograde motion of the M stars in their orbit lead to unusually short or long duration eclipses;
(ii) due to 3-body dynamics, the M-star orbit precesses with an astonishingly quick timescale of
1.74 years for the periastron (apsidal) precession, 
and 2.73 years for the inclination and nodal angle precession. 
Using the full \ik~data set, supplemented with ground-based photometry, plus 29 radial velocity measurements that span 6 years, our photodynamical modeling yields masses of
M$_{A} = 1.2713   \pm 0.0047  ~M_{\odot} \ (0.37\%)$,
M$_{B} = 0.23529 \pm 0.00062 ~M_{\odot} \ (0.26\%)$, and
M$_{C} = 0.20739 \pm 0.00055 ~M_{\odot} \ (0.27\%)$
and radii of 
R$_{A} = 1.9984 \pm 0.0027 ~R_{\odot} \ (0.14\%)$,
R$_{B} = 0.25504 \pm 0.00076 ~R_{\odot} \ (0.3\%)$, and
 R$_{C} = 0.23196 \pm 0.00069 ~R_{\odot} \ (0.3\%)$. 
We also estimate the apsidal motion constant of the M-dwarfs, a parameter that 
characterizes the internal mass distribution. 
While not particularly precise, we measure a mean apsidal motion constant, $\overline{k_{2}}$, of $ 0.046^{+0.046}_{-0.028}$,
which is approximately 2-$\sigma$ lower than the theoretical model prediction of 0.150.
We explore possible causes for this discrepancy.

\end{abstract}



\section{Introduction}\label{sec:intro}
Double-lined eclipsing binary stars are fundamental to stellar astrophysics.  
Radial velocity measurements of each star plus the photometric observations of eclipses allow us to determine the stellar masses, radii, and orbital parameters \citep{2010AARv..18...67T}. The resulting characterization of the system is independent of theories of stellar evolution.
If the binary star is actually part of a triple-star eclipsing system, then we potentially can measure the system parameters with very high precision. This is due to the additional sampling of the binary star's orbit: the eclipses caused by the third body pinpoint the relative locations and velocities of the stars. Furthermore, mutual dynamical interactions lead to eclipse timing variations, which can be measured with high precision, allowing an additional constraint on the masses and separations. 
Eclipse-timing variations in binary stars are not uncommon: In a study of 2600 eclipsing binaries in the \ik~database, 222 systems show strong evidence of  a third body in the system \citep{2016MNRAS.455.4136B}. While uncommon, 17 systems of these systems have third body eclipse events occurring with periods ranging from months to years \citep{2020MNRAS.496.4624B}.
KOI-126 (KIC 5897826) was the first triply eclipsing system of this kind to be found with the \ik~space telescope and is the most compact known \citep{2011Sci...331..562C}.   
    
\begin{figure}[ht!]
\centering
\plotone{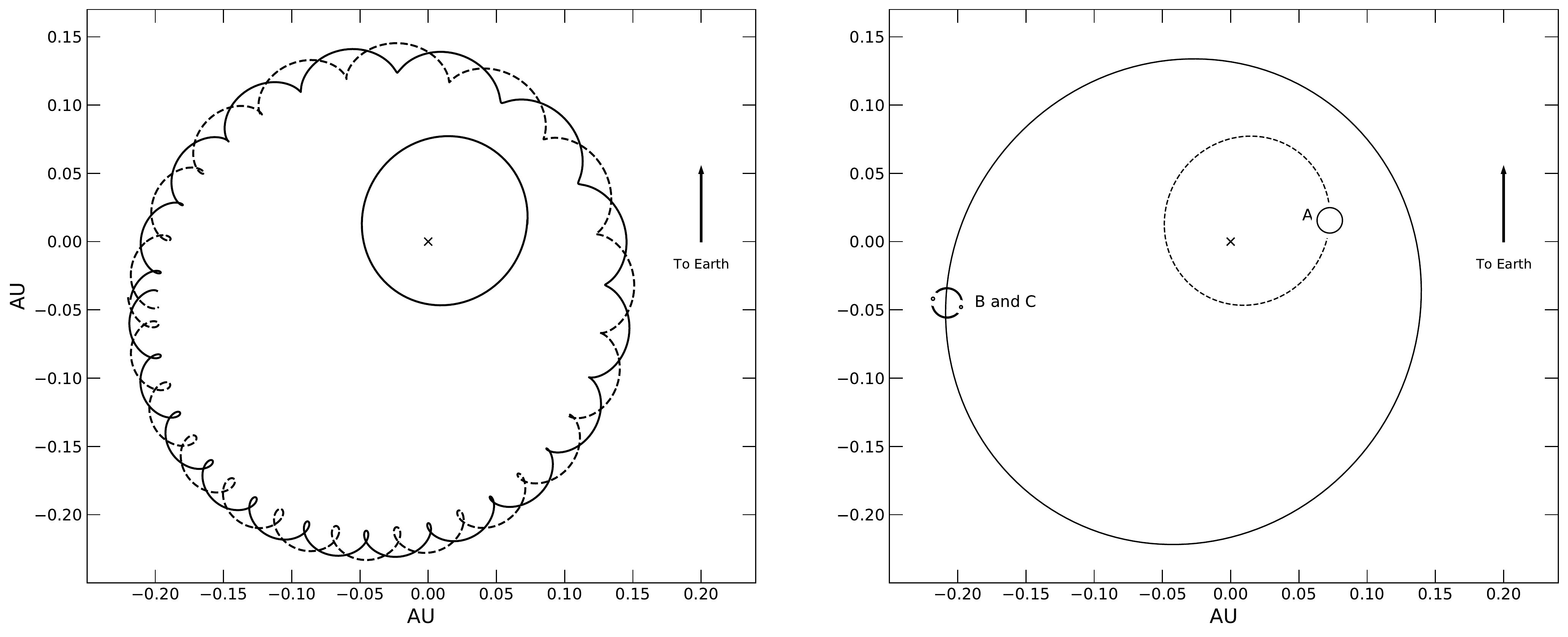}
\caption{A top-down view of KOI-126 at BJD 2455043.989. The x marks the center of mass of the system, about which each body orbits. The left panel shows the motion of each star during a single orbit around Star A. The right panel shows the scale of the system and each star with the dashed curve tracking
the approximate location of the center of mass of KOI-126 B and C while the solid curve tracks the orbit of KOI-126 A. The circles are to-scale representations of KOI-126 A, B and C. }
\label{fig:diagram}
\end{figure}

Discovered and characterized by \cite{2011Sci...331..562C}, KOI-126 is comprised of two M-stars (KOI-126 B and C) in a mutual 1.74-day orbit, which in turn is in a 33.9-day orbit about an F-star (KOI-126 A; the primary star). The inner orbit (B and C) and outer orbit (B+C around A) are aligned such that stars B+C transit star A from the perspective of Earth. The mutual eclipses of B+C are not apparent because the M stars are so faint compared to star A, and this difficulty is compounded by the precession of the B+C orbit such that the stars often do not eclipse as seen from Earth. Figure \ref{fig:diagram} shows two top down views of the system with the barycenter marked. The left panel shows the orbital motion of each body through one cycle, while the right panel shows the relative sizes of the orbits and stars.   

The motion of the stars in the B+C binary is complex, due to the binary orbiting in the gravitational potential of star A. The 3-body dynamics cause a rapid $\sim$1.7-year precession of the binary stars' line of apsides and a $\sim$2.7-year precession of the inclination and line of nodes (see \cite{2011Sci...331..562C} and our revised estimates in
Section \ref{sec:precession}).
Because the orientation of the inner stars' orbit relative to the primary star is different at each conjunction, this results in a rich, complex eclipse profile.
If the projected (plane of the sky) separation between the B+C binary is large during an eclipse of the primary (corresponding to an inner binary phase of 0.25 or 0.75), we see two well-separated eclipses of
star B and star C across star A. However, if the B+C binary is near alignment at the time of eclipse (phase 0.0 or 0.5), we observe one short eclipse superimposed over a much longer eclipse.
At these phases, the B and C stars are moving in opposite directions in their orbit, with one moving in the same direction as the orbital motion while the other is moving ``retrograde''.
This creates overlapping eclipses of very different durations.
Four of these overlapping eclipses in the \ik~dataset are also syzygy events 
in which all three stars overlap. These eclipses are of particular interest, 
carrying precise positional and relative motive information on each star.
The complex eclipses make observational data difficult to fit with a triple star model, but with this difficulty comes extremely precise stellar and orbital parameters of each body in the system when a good fit is found.   

In their discovery paper, \citet{2011Sci...331..562C} developed the ``photodynamical modeling
method'', by which we mean the equations of motion are integrated to simultaneously fit the eclipse and radial velocity data. This is necessary because the orbits do not exhibit simple Keplerian motion. The equations require terms for apsidal precession caused by tidal and rotational distortion and also general relativistic precession. Using their photodynamical model
\citet{2011Sci...331..562C} fit the the first 8 eclipse events in the \ik~data plus radial velocity data to obtain precise masses and and radii of the stars:
M$_{A}$ = 1.347 $\pm$ 0.032 M$_{\odot}$, M$_{B}$ = 0.2413 $\pm$ 0.0030 M$_{\odot}$, and M$_{C}$ = 0.2127 $\pm$ 0.0026 M$_{\odot}$ with uncertainties of 2.4\%, 1.2\%, and 1.2\%.
For the radii, they find: 
R$_{A}$ = 2.0254 $\pm$ 0.0098 R$_{\odot}$, R$_{B}$ = 0.2543 $\pm$ 0.0014  R$_{\odot}$, and R$_{C}$ = 0.2318  $\pm$ 0.0013 R$_{\odot}$ with uncertainties of 0.48\%, 0.55\%, and 0.56\%.
These are very precise values, but \citet{2011Sci...331..562C} claimed that much higher precision was possible and made the prediction that by using the full \ik~data set a precision of better than 0.1\% would be obtainable in the masses and radii.
In this paper we attempt to fulfil that prediction by utilizing the complete 4-year \ik~data, plus additional radial velocity data and ground-based eclipse observations to derive even higher precision system parameters.

\citet{2011Sci...331..562C} made an additional bold prediction: that the apsidal constant $k_2$ (also called the internal structure constant) could be measured using the full \ik~dataset to a relative precision of $\sim$1\% by measuring the precession of the orbit, ie. the shift of the argument of periastron. Analogous to the Love number or the polytropic index (or the moment of inertia or the moment of gyration), the apsidal motion constant is related to the mass distribution inside the star. For massive stars, the apsidal motion constant and asteroseismology have both provided vital constraints on their internal structure. But for low mass stars, neither of these techniques have provided any observational information. This is unfortunate, as it is the low mass stars that could benefit most from such information, e.g., the well-known radius discrepancy for low-mass stars, and the importance of M stars for exoplanet habitability considerations. In their paper examining stellar evolution models for KOI-126, \cite{2011ApJ...740L..25F} state that:
{\it  the determination of the apsidal constant will 
provide a crucial test of our stellar evolution models. 
In particular, it will test the EOS, which directly 
determines the run of density necessary for the computation 
of the apsidal motion constant. ...to accurately derive 
the apsidal motion constant to within 1\% ...will 
provide a stringent benchmark against which to test the
interior physics of low-mass stellar evolution models.}
\citet{2011Sci...331..562C} could only put an upper limit of $k_2 < 0.6$, which as they noted is not particularly useful since the models predict a value of $\sim$0.15 for stars of this mass. But with the additional data in hand, we can now attempt to precisely measure $k_2$. We note that in order to measure the apsidal precession, the eccentricity of the orbit must not be too small, or the precessional motion becomes degenerate with the orbital period. The presence of a third star will tend to inhibit orbit circularization, in theory allowing the apsidal motion to be measurable even in short-period binaries, which otherwise would tend to circularize. Thus KOI-126 appears to be 
well-suited for such investigation. \\

\subsection{Apsidal Motion Constants}\label{sec:apsidal}
Stars are not point masses, nor are they spherical. Rotation and tidal forces distort their shapes, and thus the argument of periastron (or equivalently the line of apsides) will precess. If this precession occurs on a timescale comparable to the duration of the observations, the orbits of the stars cannot be approximated as simple Keplerian motion. In the case of KOI-126 B and C, the apsidal period is a very short $\sim$1.78 years 
(\citet{2011Sci...331..562C} and see Section \ref{sec:precession}). This fast precession is caused by a combination of a third body (KOI-126 A), general relativity, and tidal forces, including rotational oblateness.

Because the orbital motion is non-Keplerian, we integrate the equations of motion to determine the positions of the three stars.
The accelerations due to three-body effects and general relativity only depend on the masses and separation between each component, treating each body in the system as a point. However, non-sphericity in the shape of each star from tidal and spin distortions produces an acceleration due to the non-zero quadropole moment and requires additional terms in the equations of motion.
We use the hierarchical Jacobian coordinate system and the acceleration equations
given in \citet{2002ApJ...573..829M} and \cite{2009ApJ...698.1778R}.
These equations contain a term known as the apsidal motion ``constant'' $k_{2}$ (equivalent to half of the Love number) which characterizes the radial mass density distribution inside the star (see \cite{2009ApJ...698.1778R} for a thorough discussion, and also \cite{2010AA...519A..57C} for a more observational-motivated discussion).
A value of $k_{2}=0$ is appropriate for a star that can be modeled as a point mass, and $k_{2}=0.75$ for a body with a uniform mass distribution. For fully convective low-mass stars like KOI-126 B and C, a value near 0.15 is expected.\\

The apsidal motion constant is determined from observations by measuring the rate of the periastron precession. This is best obtained from measurements of the primary and secondary eclipse times, which in turn gives the rate of change of the argument of periastron $\dot{\omega}$. Since there are two unknowns ($k_2$ for each star) but only one observational datum ($\dot{\omega}$), the two $k_2$ values cannot be independently determined. Thus it is common practice to measure the weighted mean value of $k_{2}$ from both stars, $\overline{k_{2}}$
(e.g.\ see \cite{2002AA...388..518C,2007IAUS..240..290G}):

\begin{equation}\label{eq:k2}
    \overline{k_{2}} = \frac{1}{c_{21}+c_{22}} \frac{P}{U},
\end{equation}
where 
\begin{equation}\label{eq:k2cij}
    c_{2i} = \left[ \left(\frac{\omega_{i}}{\overline{\omega}}\right)^{2}\left(1 + \frac{m_{3-i}}{m_{i}}\right)f(e) + \frac{15m_{3-i}}{m_{i}}g(e) \right] \left( \frac{R_{i}}{a} \right)^{5},
\end{equation}
\begin{equation}\label{eq:fe}
    f(e)=(1-e^{2})^{-2},
\end{equation}
and
\begin{equation}\label{eq:ge}
    g(e)= \frac{1 + \frac{3}{2}e^{2} + \frac{1}{8}e^{4}}{(1-e^{2})^{5}}.
\end{equation}
In these equations, $c_{2i}$ are the weights used to calculate $\overline{k_{2}}$, $P$ is the period of the binary, $U$ is the apsidal-motion period (time for one turn of the apside), 
$a$ is the semi-major axis, $e$ is the eccentricity, $m_{i}$ is the stellar mass of 
the i$^{th}$  star, $R_{i}$ is the radius of the i$^{th}$ star,
and $\frac{\omega_{i}}{\overline{\omega}}$ is the ratio
between the rotational angular velocity of the i$^{th}$ 
star and the average orbital angular velocity.
The functions $f$ and $g$ are only functions of the eccentricity.
For KOI-126 B and C we find that there is a negligible difference between 
the weighted $\overline{k_{2}}$ and their unweighted mean. For simplicity $\overline{k_{2}}$ will refer to the unweighted mean of $k_{2,B}$ and $k_{2,C}$ for the rest of this work.
The apsidal-motion period, $U$, is related to the precession by
\begin{equation}\label{eq:k2period}
    U = \frac{360^{\circ} P }{\dot{\omega}_{obs}},
\end{equation}
where $\dot{\omega}_{obs}$ is the observed rate of apsidal precession in units of degrees/cycle and P is in units of days/cycle. 
The apsidal motion period can be directly measured from the period of a cycle in an eclipse timing O-C diagram, after other effects (e.g.\ light travel time) are accounted for. However, often a full precession cycle is not observed, and $U$ must be inferred from the shape of the O-C curve. For KOI-126, the extremely complex nature of the eclipses renders both the measurement and the interpretation of an O-C diagram useless for measuring the apsidal motion. Fortunately we can constrain $k_2$ directly as a free parameter in our photodynamical modeling, i.e., instead of using the mid-eclipse times, we use the full eclipse shape (duration, depth, and timing).

Along with asteroseismology, measuring the apsidal motion constants are currently 
the only methods of observationally constraining the internal density structure of high mass stars, making it an invaluable tool for testing the theory of stellar interiors. 
In the case of KOI-126 B and C, asteroseismology is not possible since such M stars
are not known to pulsate, the stars are intrinsically faint, and the light from KOI-126 A dilutes their signal by a factor of $\sim$1,400. However, it should be noted that asteroseismology of KOI-126 A is very challenging but possible, and a comparison of the results of such a study with our photodynamically-derived system parameters would be extremely valuable.

While asteroseismology is not feasible for M stars, measuring the apsidal motion constant is in principle possible, if the binary star conditions are favorable. Primary and secondary eclipses must be present, the period must be short, and the orbit must be eccentric enough for the precession to be measurable on a human timescale. The latter two requirements tend to be mutually exclusive, as short orbital periods are usually associated with zero eccentricity (due to tidal circularization).
For example, the well-known M-star binary system CM Dra has a very small eccentricity
(e = 0.0051), and so its precession period is $\sim$5435 years \citep{2010AARv..18...67T}.
KOI-126 B and C are similar to CM Dra A and B in terms of masses, radii, period, and eccentricity, however the orbit of KOI-126 B and C is precessing remarkably fast with a 
total apsidal period of only 1.741 years (see Figure \ref{fig:inner}).
This enhanced precession rate is driven primarily by the third body, KOI-126 A, but could potentially make the values of $k_{2}$ much more readily measurable. 

\section{Data}\label{sec:data}

\subsection{\ik~Light Curves}\label{sec:Kepler}
KIC 5897826 was observed by \ik~in Quarters 0, 1, and 2 in 30-minute long cadence mode before being flagged as an object of interest due to its unusual eclipse profiles (at one point it was even thought to possibly be a binary planet) \citep{2010Sci...327..977B}. Designated as KOI-126, subsequent observations were done in 1-minute short cadence mode, except for Quarter 8 for which only long cadence data are available.

\begin{figure}[ht!]
\centering
\plotone{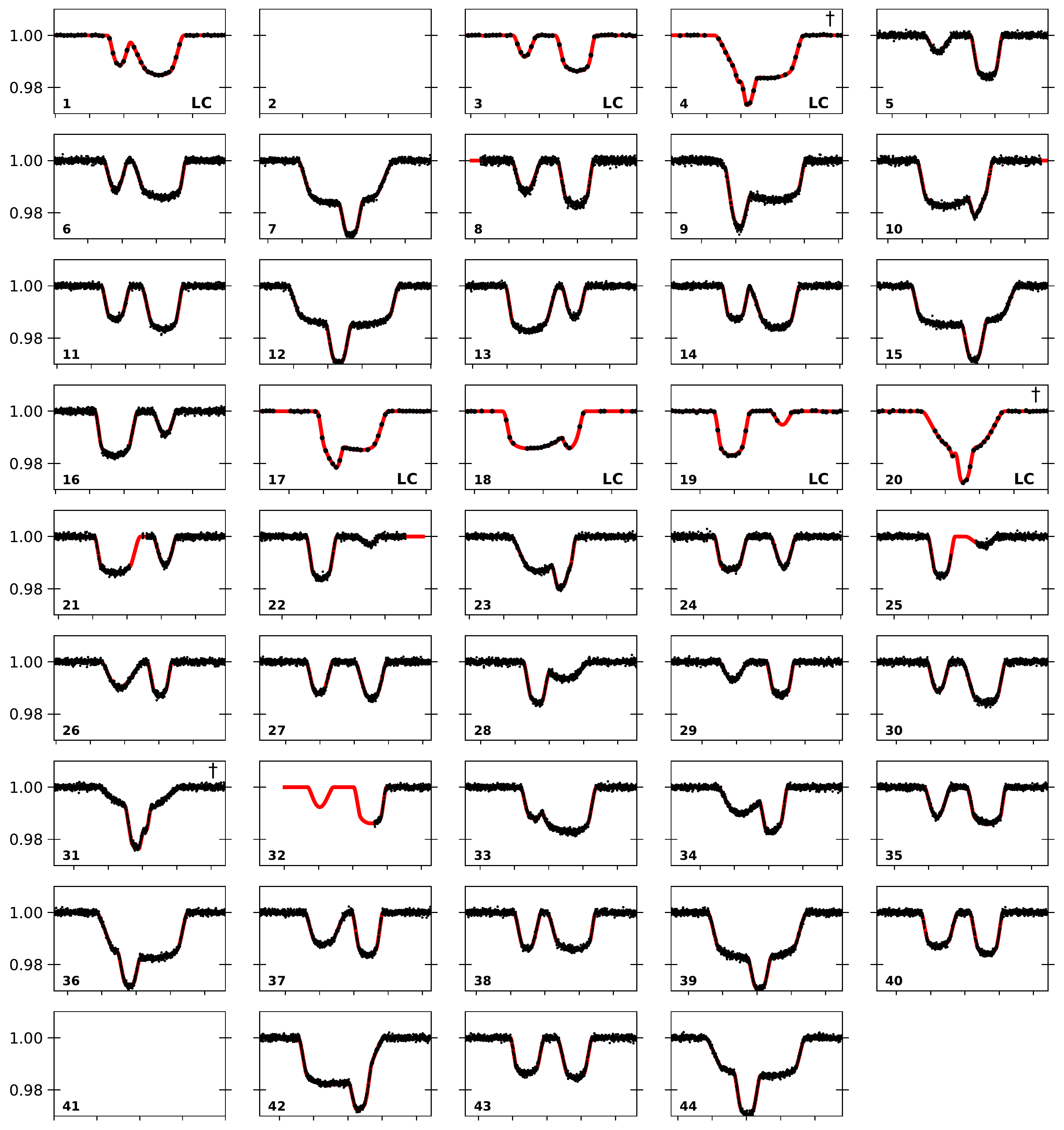}
\caption{Each eclipse of the inner orbit across KOI-126 A and the best fitting model observed by the \ik~space telescope. The eclipse number is given in the lower left hand corner of each plot. Eclipses 1, 3, 4, 17, 18, 19, and 20 only have long cadence data available. Eclipse numbers 2 and 41 occurred during a gap in the data, so was not observed. The width of each plot is 1 day, a substantial fraction of the 1.72 day orbital period of the KOI-126 B+C pair. Daggers indicate syzygy events in which KOI-126 A, B, and C 
eclipse each other simultaneously. Eclipses 5 through 12 were used by \citet{2011Sci...331..562C} }
 \label{fig:kepler}
\end{figure}

\begin{figure}[ht!]
\centering
\plotone{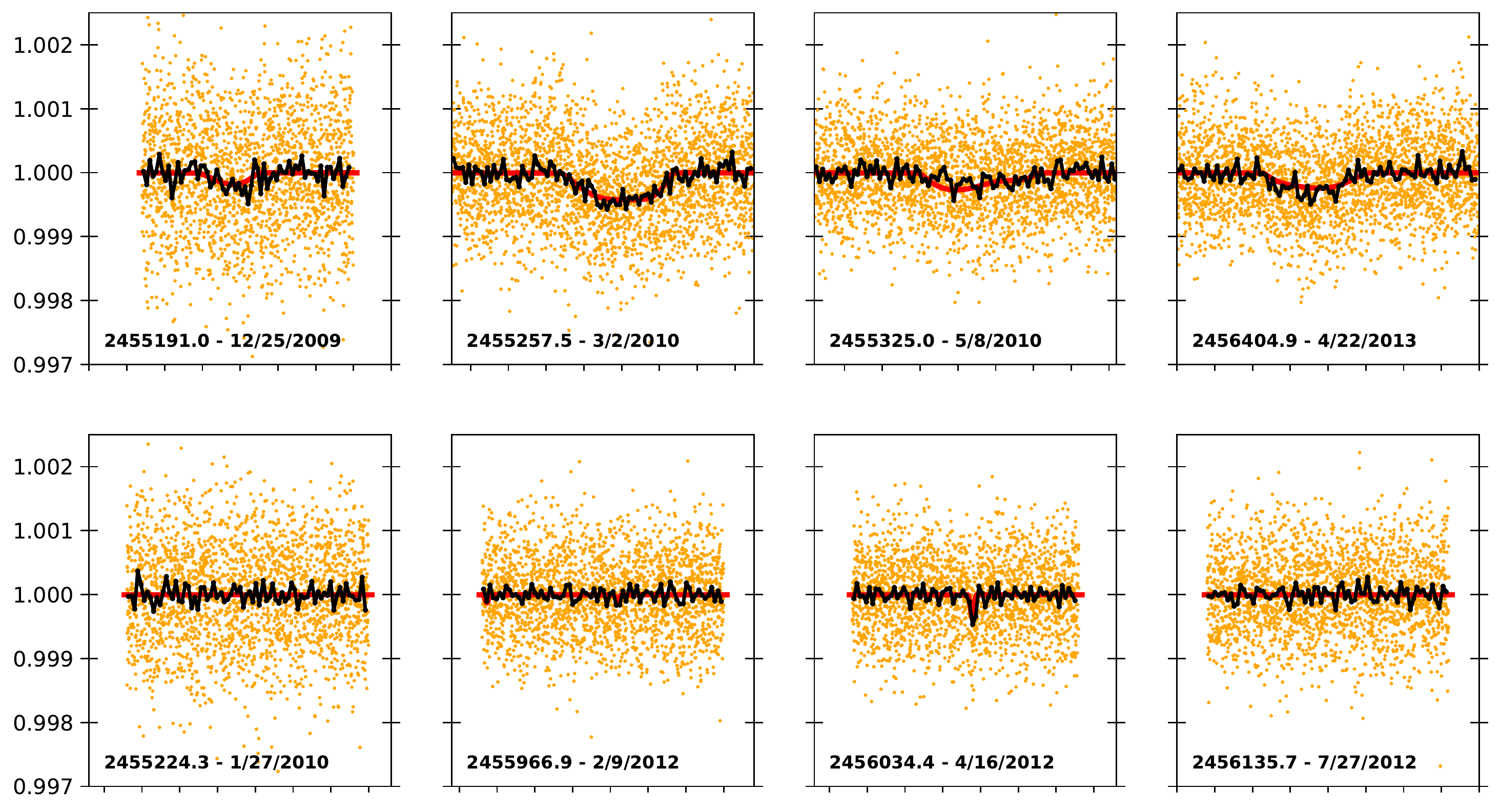}
\caption{Examples of occultations (secondary eclipses) of Stars B and C behind Star A. The orange points are the \ik~short cadence data, the black line shows the data binned to 30 minutes, matching the \ik~long cadence data, and the red line shows the best fitting model. Each plot shows the starting BJD, the corresponding UTC date, and has a duration of 2 days. The top row shows the visible occultations present in the \ik \ data, while the bottom row shows a number of near misses and eclipses of the inner binary.}
 \label{fig:kepler-occul}
\end{figure}

All available data were retrieved from the MAST archive in 2017 corresponding to \ik~Data Release 25. To remove any systematic calibration errors the Simple Aperture Photometry (SAP) calibrated data were detrended in segments. A 5\textsuperscript{th} order polynomial was fit to segments of the out-of-eclipse and out-of-occultation light curve, then each segment was divided by its polynomial fit, and then the segments were recombined to form the light curve. 
Figure \ref{fig:kepler} shows all of the \ik~eclipse data used in this study, noting that eclipses number 2 and 41 were not observed. In addition to the eclipses, occultations  (Star B or C behind Star A) can be informative, despite being shallow and barely detectable. \cite{2011Sci...331..562C} fit for two such events, but every eclipse event has a potential corresponding occultation event, so sections of the light curve when an occultation could potentially occur are included in the data. A majority of these possible occultation events do not exhibit any visible occultation, partly because of noise, but also because the precession causes the stars to be misaligned and not produce an occultation at each conjunction.
At such times, the flat occultation data are still helpful, as they can penalize models which have occultations when there should be none. 
Figure \ref{fig:kepler-occul} shows the clearly visible occultations in the \ik~dataset and a number of near misses.

Since out-of-eclipse and out-of-occultation data do not provide much useful information, any data not within a window of five times the eclipse duration centered on mid-eclipse are removed. This speeds up computation, but more importantly, it prevents the uninformative flat parts of the light curve from dominating the goodness of fit merit function (chi-square or likelihood). Note that while KOI-126~B and C do eclipse each other frequently, these eclipses are extremely shallow (see the bottom plots in Figure \ref{fig:kepler-occul}) due to the dilution of light by KOI-126~A.
In theory the individual eclipses could be combined to increase the signal, but this is problematic. The orbit precesses so quickly that combining the eclipses is not valid: the inclination changes from eclipse to eclipse, even to the point that the stars no longer eclipse each other. Since this effect, plus the light-travel time delay due to the orbital motion, cannot be accounted for in a {\em model-independent} way, we do not include the B+C eclipses in our final data set.   

In the discovery paper by \citet{2011Sci...331..562C}, 247 days of short cadence data containing eight full eclipse events (stars B+C eclipsing star A) were available. In this work we include additional observations made by \ik, which now span a total of 1457 days and contain 42 eclipses and 40 potential occultations.   
This notably larger amount of \ik~data will allow us to significantly improve the precision of the system parameters.
We examined TESS data on KOI-126 using the \texttt{eleanor} python package and they are not particularly helpful because of the low signal-to-noise ratio and systematics, but they may be of value in the future if an improved calibration is available \citep{2015JATIS...1a4003R,2019PASP..131i4502F}.

\begin{figure}[ht!]
\centering
\plotone{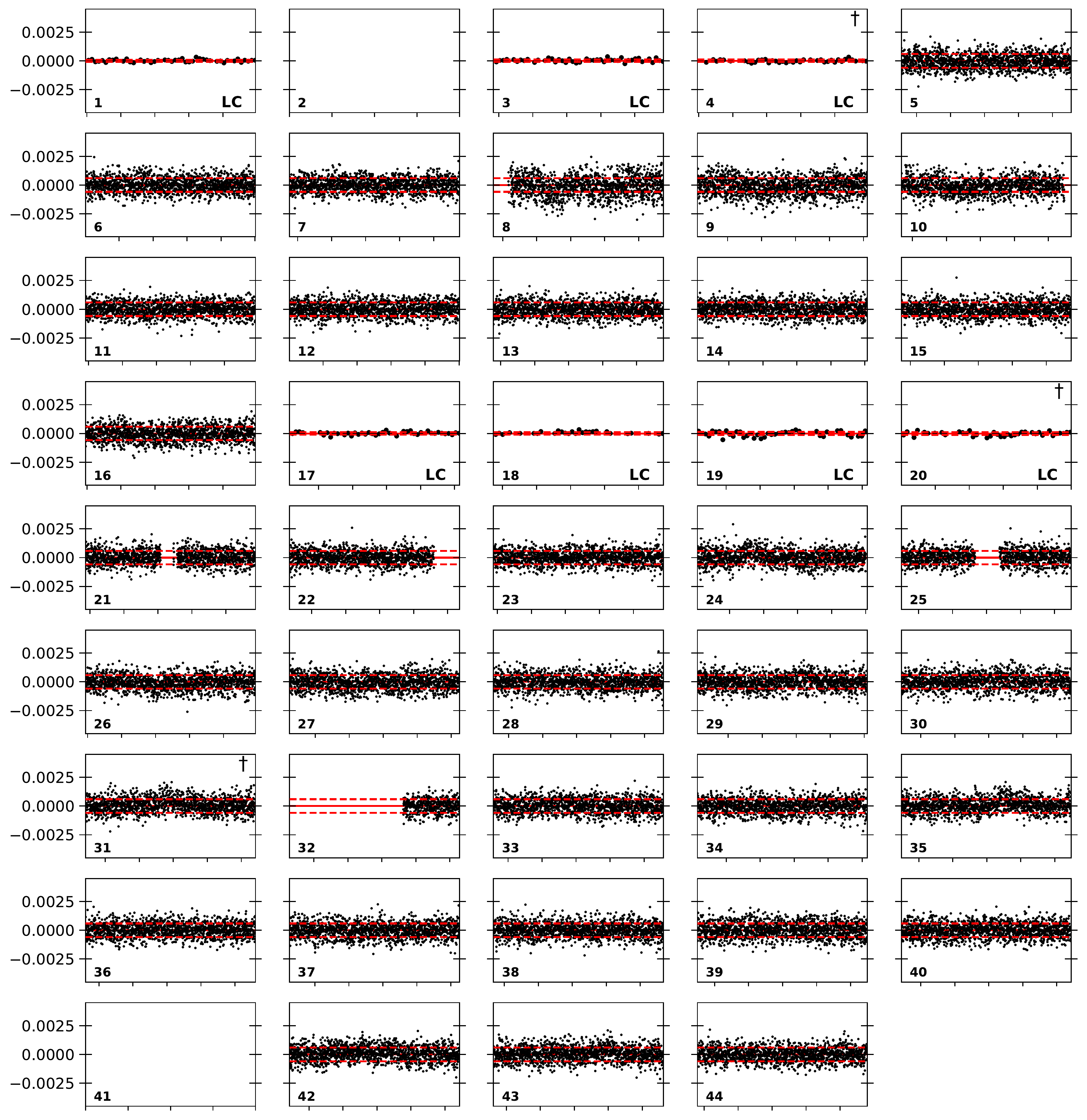}
\caption{The residuals corresponding to Figure \ref{fig:kepler} using the best fit model. The dashed red lines indicate the 1-sigma range for the median \ik~uncertainty, $\sim$33\% of the points fall outside these bounds.}
 \label{fig:kepler-res}
\end{figure}

\subsection{Mount Laguna Observatory Photometry}\label{sec:MLO}
\begin{figure}[ht!]
\centering
\plotone{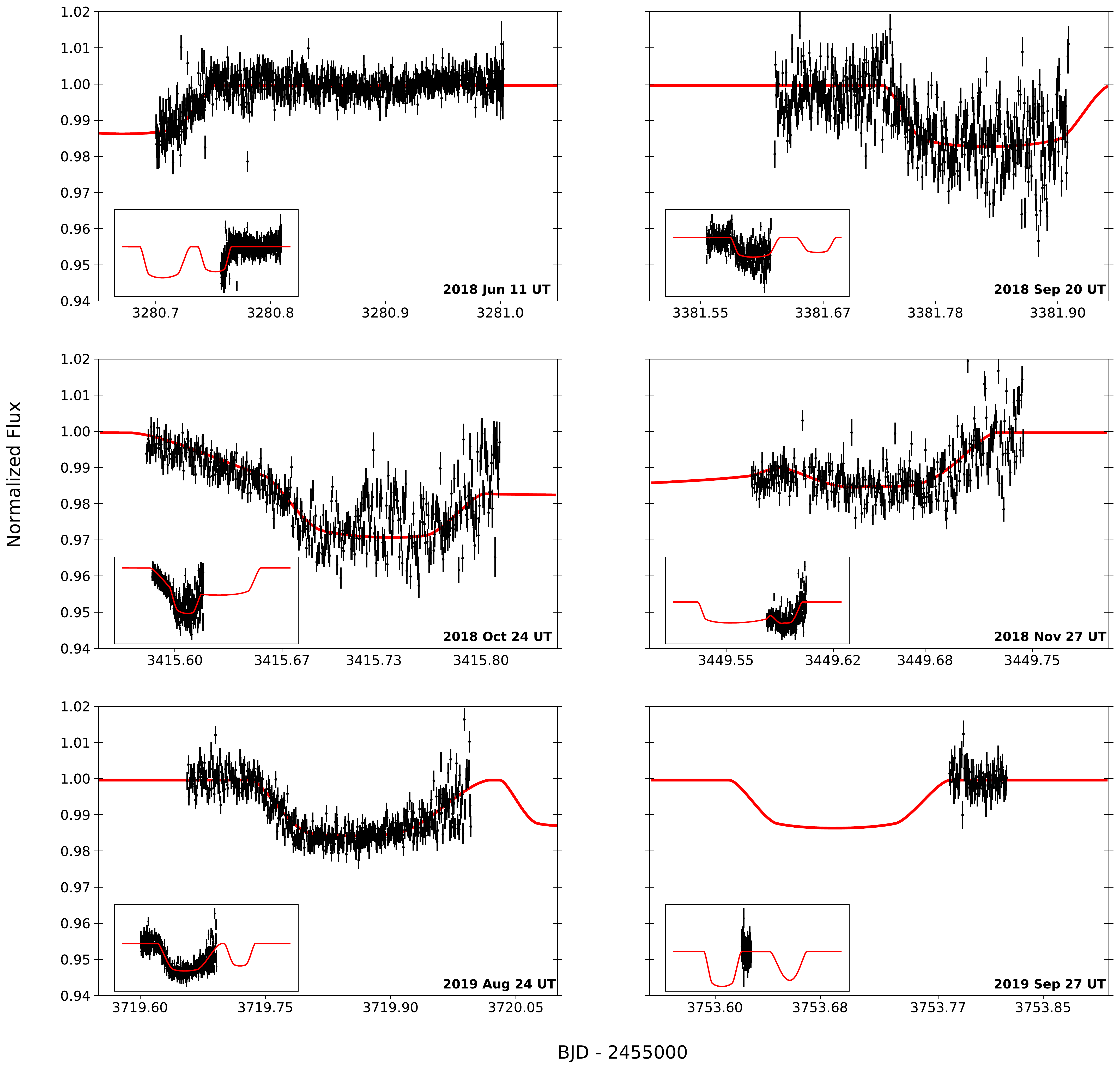}
\caption{Partial eclipses of KOI-126 observed from the 1-meter Mount Laguna Telescope 
in the Johnson-Cousin's R-band.
Each inset window has a length of 18 hours showing the full eclipse profile.
}
\label{fig:mlo}
\end{figure}

A ``typical'' KOI-126 eclipse lasts for approximately 12 hours, making ground observations of the full eclipse difficult. But partial eclipses, especially ones that catch egress, ingress, or mid-eclipse are still very useful as they provide important timing constraints. Partial eclipse events for KOI-126 were observed during the summers of 2018 and 2019 using the Mount Laguna Observatory (MLO) 1-meter telescope in the Johnson-Cousins R-band.  These data and their epoch are shown in Figure \ref{fig:mlo}. 
Exposures of 45-60 seconds were used to maximize temporal resolution while still giving reasonable signal-to-noise ratio. A faint star, 2MASS	19495481+4106506, (Gaia mag = 16) unresolved in the \ik~data is located about 8 arcseconds from KOI-126. With a pixel scale of 0.4 arcseconds/pix and an average seeing of 2-3 arcsecond during our observations, these stars remained well separated for each of our observations and the aperture size was set to contain only KOI-126.

Standard bias and flat field calibration and aperture photometry was performed using the AstroImageJ code (AIJ; \citet{2017AJ....153...77C}). Observation times were converted from UT to BJD using the UTC2BJD calculator in the AIJ code. For each night, differential photometry was carried out using the same 7 comparison stars within 3 arcminutes of KOI-126. 2MASS 19495481+4106506 was examined to ensure it is not contributing to the eclipse profile variability seen in the \ik~data. Its total integrated flux is $\approx 5$\% of the total flux from KOI-126, which is fit for in the \ik~data as contamination.
Our modeling confirm it adds $\sim$ 1-4\% additional light in the Kepler bandpass.
For the MLO data the stars are sufficiently resolved such that negligible contamination is present.
Due to poor weather earlier in the night the final MLO observation missed egress, however it is sufficiently close to egress that it could act as a constraint on the egress timing. The inclusion of these MLO eclipses taken 5 years after the primary \ik~mission ended extends the range of photometric data to $\sim$10 years (3719 days), a $\sim$14 times larger time span than the 
\citet{2011Sci...331..562C} study.

\subsection{Spectroscopy and Radial Velocities}\label{sec:RV}
Twenty-nine spectra of KOI-126 were obtained and used to form our set of radial velocity measurements. KOI-126 A is about 1400 times brighter than KOI-126 B and C combined, so the spectra are single-lined.
Twenty-three of the spectra were taken using the Tillinghast Reflector Echelle Spectrograph (TRES) on the 1.5 meter Tillinghast Reflector from the Fred L.\ Whipple Observatory in Mt. Hopkins, AZ \citep{gaborthesis,TRES}. TRES has a resolving power of R=44,000 with a wavelength range of 3850-9090 \r A. The other six spectra were taken at the McDonald Observatory in Texas using the Tull Coude Spectrograph on the 2.7 meter Harlan J. Smith telescope. Using the medium fiber, these spectra have a resolving power of R=60,000 
with a wavelength range of 4580 to 6520 \r A. Precise radial velocity measurements for each of these spectra were obtained using multi-order cross-correlations, after spectral orders containing strong atmospheric lines, low signal-to-noise ratio, and known reduction problems (mostly in the red regions) were omitted \citep{2010ApJ...720.1118B}. The useful wavelength ranges were 4580-6562 \r A from TRES and 4500-6680 \r A from the McDonald / Tull spectrographs. The radial velocities are given in Table \ref{tab:RV-data}; these have not had a systemic gamma-velocity offset correction applied.
The systemic velocity of KOI-126 and the different spectrograph zero-point offsets are of course values that need to be determined when fitting the radial velocity data, but these are not treated as free parameters in our MCMC modeling (described below). Rather, these additive velocity terms can be independently exactly optimized for a given model radial velocity curve in one internal iteration (i.e.\ just one shift is needed per radial velocity data set). The combined systemic and zero-point offsets for the two spectrographs are given in Table \ref{tab:RV-data}.
Note that the radial velocities reported in Table \ref{tab:RV-data} slightly differ from those in \citet{2011Sci...331..562C}. The spectra were re-measured and a correction was made after a small artificial trend in the zero-point velocity for the TRES observations was found (caused by two of the TRES RV standard stars having companion stars). In addition, the velocities were shifted to be on the IAU absolute radial velocity scale. An additional 0.1 \kms \ should be added in quadrature to the uncertainties if absolute radial velocities are required.

\begin{figure}[ht!]
\centering
\plotone{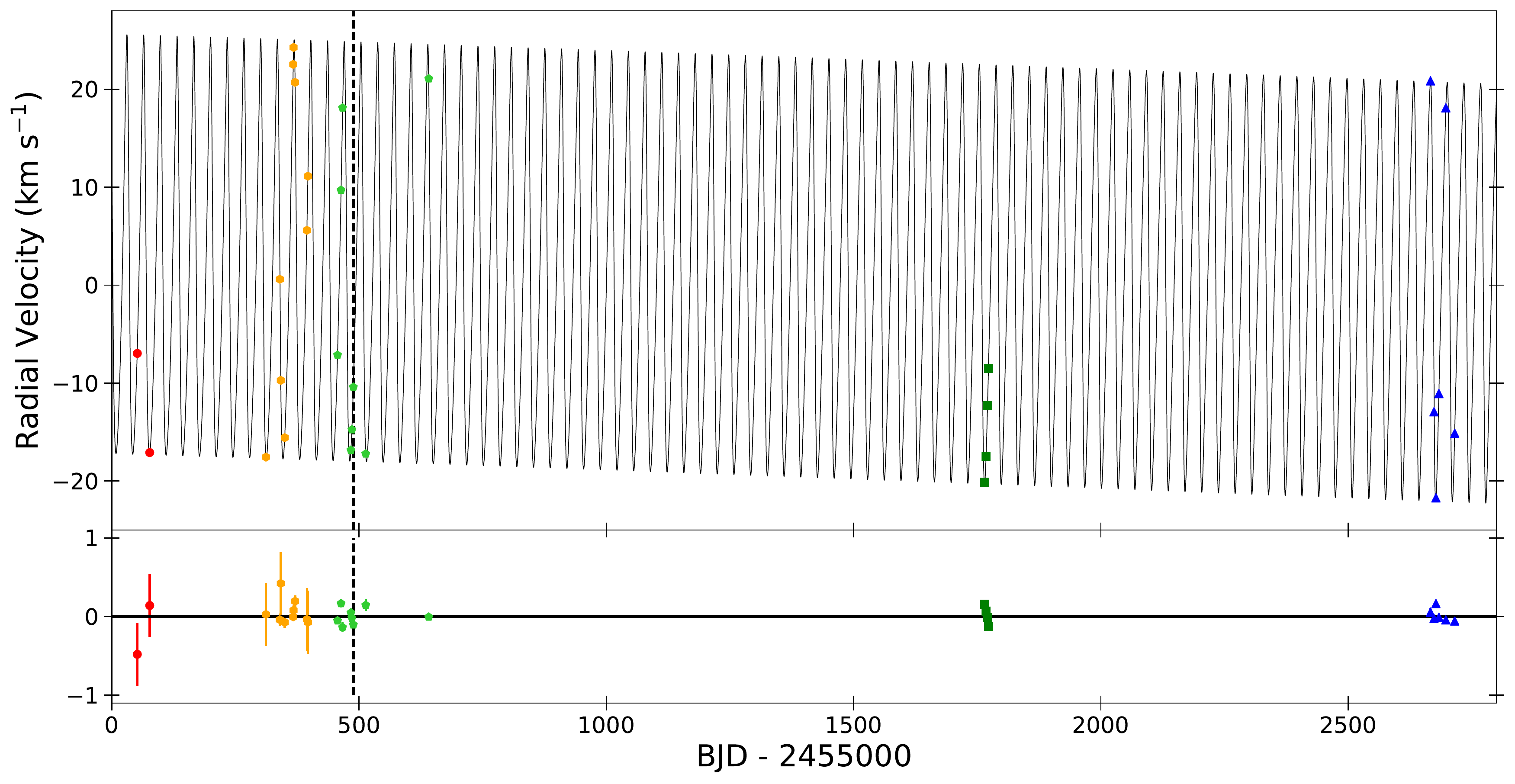}
\caption{The full span on the KOI-126 A radial velocity data and best fitting model. 
The bottom plot shows the residuals.
The color and shape of each data point is assigned based on their observation epoch which 
can be seen in each subplot of Figure \ref{fig:RVzoom}. The data before the vertical dashed line 
were used by \citet{2011Sci...331..562C}, while points after are new to this work.
}
\label{fig:RVfull}
\end{figure}

\begin{figure}[ht!]
\centering
\plotone{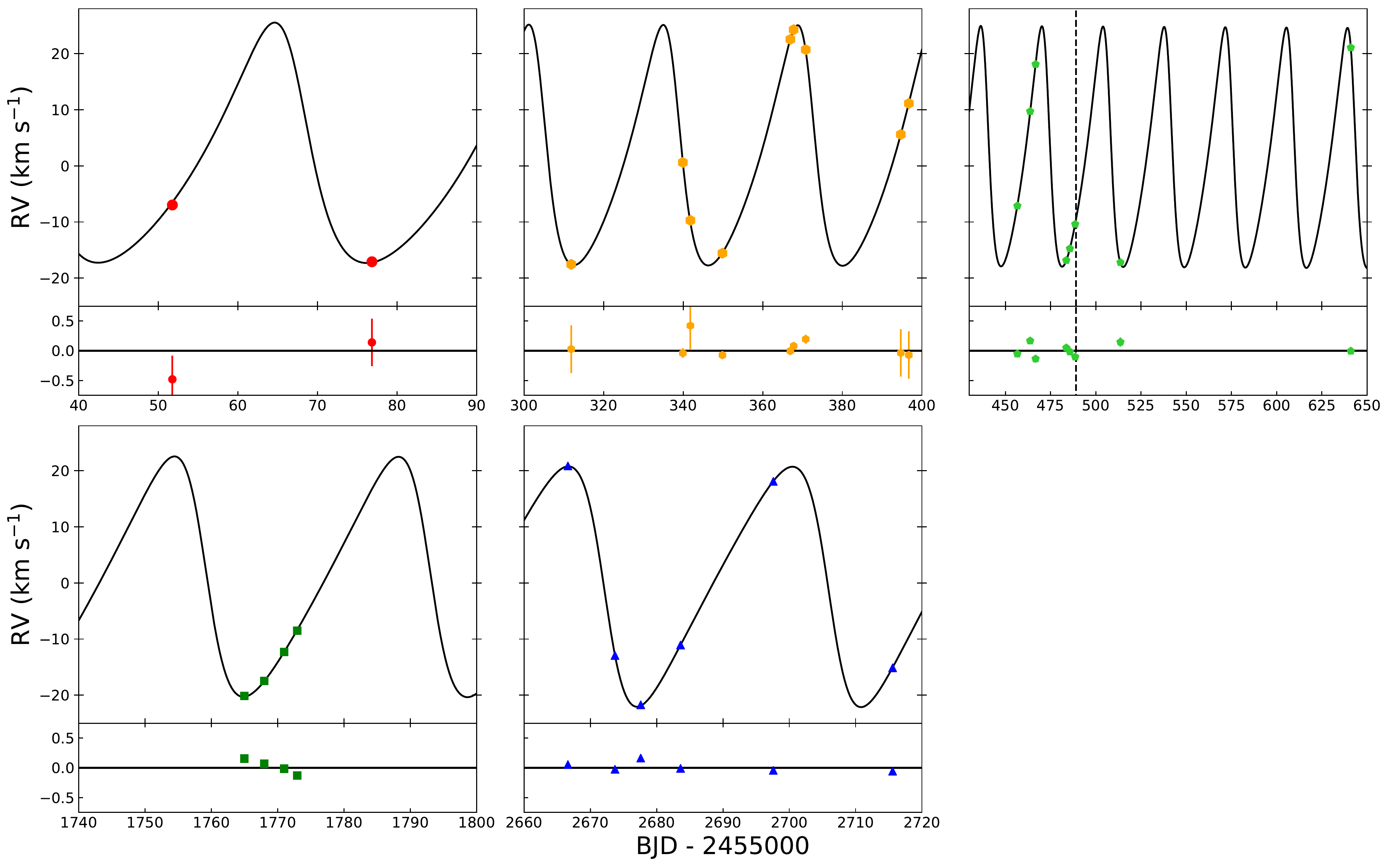}
\caption{Similar to Figure \ref{fig:RVfull}, but showing only windows around the observation epochs.}
 \label{fig:RVzoom}
\end{figure}

\begin{figure}[ht!]
\centering
\plotone{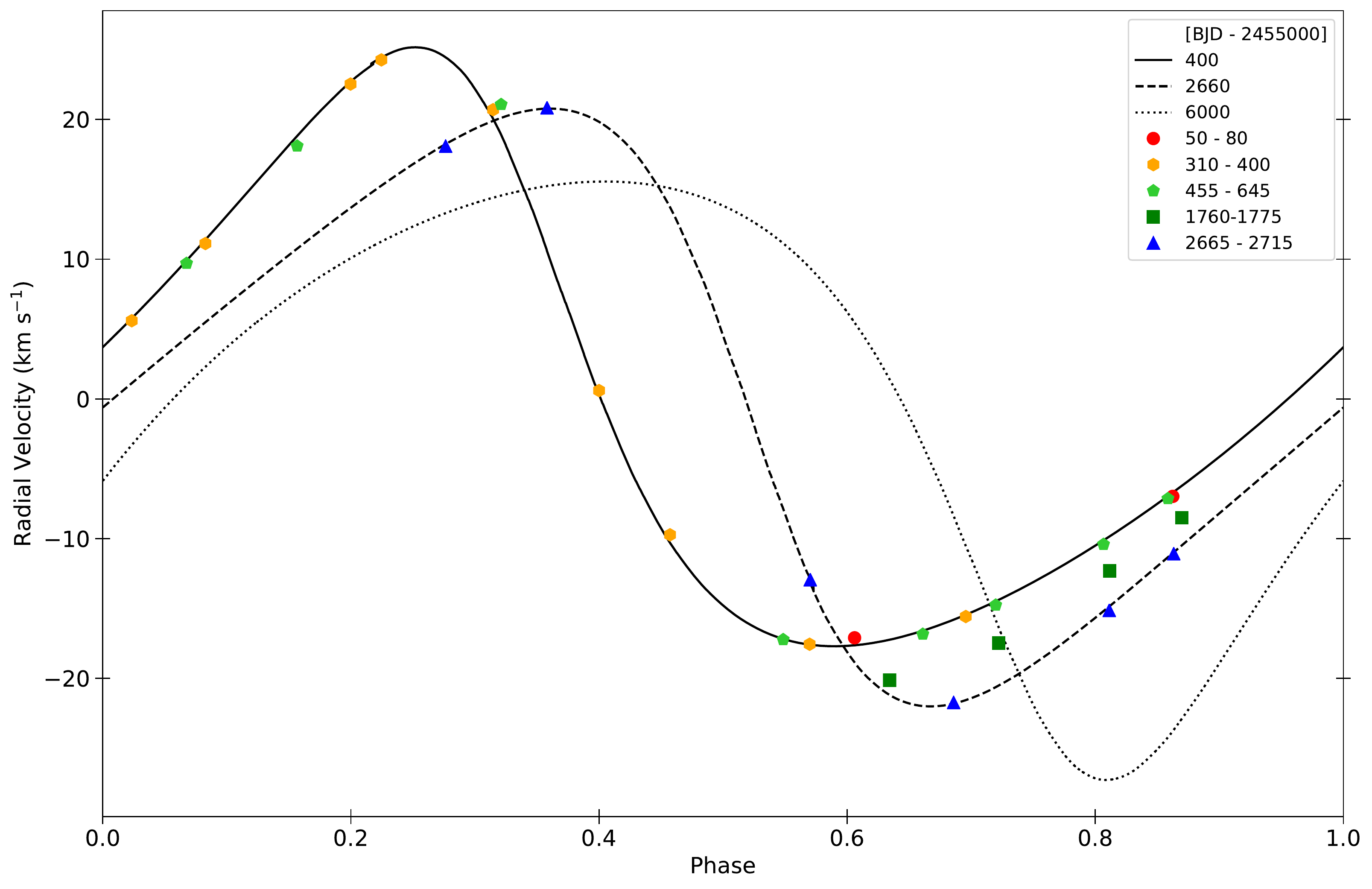}
\caption{The folded radial velocity data and model for KOI-126 A. 
Due to the ever-changing nature of system the phase of each data point and model were folded using 
the instantaneous periods and times of conjunction generated from a dynamical integration of the system. The legend gives the times corresponding to each folded RV model curve and data point in BJD - 2455000. The solid and dashed black lines shows the radial velocity curve near the start and end of the data set respectively, while the dotted line shows what it is expected to look like in 2025, around 9 years after the latest RV observation. The color and shape of each data point correspond to the observation windows seen in Figure \ref{fig:RVzoom}.}
\label{fig:RVfold-final}
\end{figure}

At the time of the \citet{2011Sci...331..562C} study, the six McDonald observatory spectra and ten of the TRES spectra were available.  In this work we make use of an additional thirteen TRES radial velocity measurements which increases the temporal baseline of radial velocity data by 6 years. Figures \ref{fig:RVfull}, \ref{fig:RVzoom}, and \ref{fig:RVfold-final} show the radial velocities and best fit model for KOI-126 A.  Note that due to the rapid precession of the orbits caused by 3-body interactions (discussed in Section \ref{sec:methods} and shown in Figure \ref{fig:outer}), we cannot simply use a single period to phase fold the observations; the period and argument of periastron change significantly over the span of our observations. The radial velocities are instead fit in time, as shown in Figure \ref{fig:RVfull}, and to show the phase-folded curves in Figure \ref{fig:RVzoom} we fold the model and data using the instantaneous (osculating) period and conjunction times generated from a dynamical integration of the system.

Figure \ref{fig:RVfold-final} illustrates the orbital evolution well: near the start of the observations the radial velocity curve is the solid black curve, and near the end the shape has changed to the dashed curve. The curve will continue to evolve into the dotted curve in 2025, 8 years after the final RV data were taken. All the observations fall between the solid and dashed curves, as expected. The apparent shift in the zero-point velocity seen in this figure, and even more clearly in Figure \ref{fig:RVfull} is a consequence of the changing orientation of the system as viewed from Earth. At these times there is a downward (negative) trend in the systemic velocity, but in the future it will reverse. This cyclical behaviour has a period of 58.86 years. 

\begin{deluxetable}{crcrc}
\tablecaption{Radial velocities of KOI-126 A \label{tab:RV-data}}
\tablewidth{0pt}
\tablehead{
\colhead{BJD} &
\colhead{RV$_1$} &
\colhead{uncertainty} &
\colhead{Telescope} \\
\colhead{(-2,455,000)} &
\colhead{(km s$^{-1}$)} &
\colhead{(km s$^{-1}$)} &
\colhead{}
}
\startdata
51.75391  & -34.797  & 0.400 & Tull \\
76.82813  & -44.921  & 0.400 & Tull \\
311.85938 & -45.383  & 0.400 & Tull \\
339.90171 & -27.243 &  0.080 & TRES 0 \\
341.81250 & -37.550  & 0.400 & Tull \\
349.86223 & -43.415 &  0.071 & TRES 0 \\
366.92702 & -5.309 &  0.054 & TRES 0 \\
367.76750 & -3.579 &  0.061 & TRES 0 \\
370.79497 & -7.143 &  0.076 & TRES 0 \\
394.70313 & -22.230 & 0.400 & Tull \\
396.71484 &  -16.705  & 0.400 & Tull \\
456.62049 & -34.975  & 0.045 & TRES 1 \\
463.69427 & -18.137  & 0.051  & TRES 1 \\
466.71428 & -9.740  & 0.062 & TRES 1 \\
483.68109 & -44.678  & 0.042 & TRES 1 \\
485.66898 & -42.595  & 0.048 & TRES 1 \\
488.61028 &  -38.248  & 0.050 & TRES 1 \\
513.62608 & -45.061  & 0.075& TRES 1 \\
641.00276 &  -6.769  & 0.024 & TRES 1 \\
1764.96562 & -47.973 & 0.044  & TRES 2 \\
1767.94373 &  -45.308 &  0.042 & TRES 2 \\
1770.96550 &  -40.146 &  0.051  & TRES 2 \\
1772.93731 & -36.335  &  0.042 & TRES 2 \\
2666.61949 &  -7.012  & 0.054 & TRES 3 \\
2673.71745 &  -40.786  & 0.030  & TRES 3 \\
2677.60059 &  -49.575  & 0.053 & TRES 3 \\
2683.60343 &  -38.926  & 0.036 & TRES 3 \\
2697.58798 & -9.760 &  0.044 & TRES 3 \\
2715.58030 & -42.980 &  0.026 & TRES 3 \\
\enddata
\tablenotetext{}{
The combined systematic and zero-point offsets are $-27.8411 \pm 0.0027$ m/s and $-27.8289 \pm 0.0037$ m/s for the TRES and TULL radial velocity data respectively.}
\end{deluxetable}
                
\section{Methods}\label{sec:methods}

To simultaneously fit the \ik, MLO, and radial velocity data we employ 
the Eclipsing Light Curve (ELC) code, a photodynamical model capable of tracking the dynamics and fluxes from multi-body stellar and exoplanet systems
\citep{2000AA...364..265O,2012Sci...337.1511O,2019AJ....157..174O}. 
Numerical integration is used to solve the Newtonian equations of motion for the three bodies, with additional terms included to account for general relativistic and tidal precession, as described in Section \ref{sec:apsidal}.

To test the tidal contribution from KOI-126 A on B and C we simulate a binary system containing a star identical to KOI-126 A and another with the same mass as KOI-126 B and C combined and radius equal to their semi-major axis. A value for $k_{2}$ was approximated using a star of similar mass for which
$k_{2}$ has been observationally measured (IT Cas: $M = 1.3315 M_{\odot}, k_{2}
= 0.0038$ \citep{2001AstL...27..712K}), while 0 was used for the B and C stand-in. 
The resulting binary has a total tidal and rotational precession rate of  
$ \dot{\omega} = 6.8 \times 10^{-11}$ deg/cycles, while the GR contribution
is $ \dot{\omega} = 8.6 \times 10^{-5}$. With a difference of 6 orders of magnitude 
the tidal contribution to the precession rate from KOI-126 A on B and C is 
negligible. So only the tidal interactions between KOI-126 B and C are included, 
with $k_{2,B}$ and $k_{2,C}$ allowed to be free parameters. The effects of 
GR between all of the stars is included.

The system is modeled in a hierarchical (Jacobian) coordinate system 
such that positions are relative to the center of mass of the inner binary 
(KOI-126 B and C). The integration is then performed using a 
12\textsuperscript{th} order Gaussian Runge-Kutta symplectic integration scheme 
in Cartesian coordinates. The origin of the coordinate system is the center 
of mass but due to numerical round-off error this will tend to wander; this gives us a handle on the precision of our numerical integration. At the end of the 3755 day integration the center of mass has moved by less than one meter, which is completely negligible. The starting positions, velocities, masses, and instantaneous Keplerian orbital parameters at the reference epoch are given in Table \ref{tab:dynamics} 
in high precision to allow our work to be reproduced and/or independently confirmed.

As seen in Figure \ref{fig:kepler} the eclipse profiles of multiple overlapping bodies have complicated shapes. To model these profiles we use the method described by \citet{2018AJ....156..297S}
which allows for the efficient computation of light curves for any number of simultaneously eclipsing and overlapping spherical bodies.The syzygy events are especially useful as they give very precise locations and velocities for all three bodies. Four such events occurred during the primary \ik~mission and are denoted by daggers in Figure \ref{fig:kepler} and \ref{fig:kepler-res}.   

The parameter space of KOI-126 is complex, which motivated the use of multiple optimization techniques to sufficiently explore the parameter space to find the best-fit solution and the confidence intervals for each parameter. We started with the parameters in \citet{2011Sci...331..562C} and set wide boundaries to allow the optimization methods to sufficiently wander.
A combination of Differential Evolution Markov Chain Monte Carlo (DEMCMC) \citep{2006S&C....16..239T},
Nested Sampling \citep{2004AIPC..735..395S}, and a modification of an adaptive amoeba direct search algorithm \citep{10.1093/comjnl/7.4.308,adaptamoeba} were employed. 
The final run was done using DEMCMC to sample the posterior distribution 
of each parameter and estimate their uncertainties. 

Our final dataset is based on the converged populations of 20 DEMCMC runs done in parallel, each made up of 192 chains. Each was allowed to sample parameter space for around 11,500 generations with burn in achieved for all runs after 200 generations , after which the auto-correlation period was $\sim$100 generations for each run. For our final posterior sample we start at generation 1,500 and sample every 200th generation from each chain to reduce inter-generational correlations.

The final best-fit model has a $\chi^{2}$ = 147974 with 139391 degrees of freedom and 30 fitting parameters as shown in Table \ref{tab:final-parms}, which results in a {$\chi^{2}_{red}$ = 1.062}. Because a majority of the data are contained in the \ik~dataset the $\chi^{2}$ is dominated by the \ik~photometry. To show the quality of the fit for each data set we compute the RMS of the residuals and compare it to the median uncertainty of each data set. The \ik~data has 137219 points with a median uncertainty of 0.00059 and RMS of 0.00061 in units of normalized flux; this is 98.4\% of all the measurements. The MLO data has 2202 points with a median uncertainty of 0.0029 and RMS of 0.0053. The TRES RV data has 23 points with a median uncertainty of 0.050 $km~s^{-1}$ and RMS of 0.097, and the TULL RV data has 6 data points with a median uncertainty of 0.4 $km~s^{-1}$ and RMS of 0.27.
We deem this as an acceptable fit considering there are systematic noise terms not accounted for during data calibration or modeling (e.g. effects caused by star spots, faculae, stellar pulsations, as well as instrumental effects) and we do not attempt to boost the RV uncertainties to account for stellar jitter. 
The best-fitting model, data, and corresponding residuals to the \ik~data are shown in Figures \ref{fig:kepler} and \ref{fig:kepler-res}. Figure \ref{fig:mlo} shows the R-band model fit with the MLO data. The radial velocity data, best fit model, and residuals were shown in Figures \ref{fig:RVfull}, \ref{fig:RVzoom}, and \ref{fig:RVfold-final}.  

\begin{deluxetable}{rrrr}
\tablecolumns{4}
\tabletypesize{\footnotesize}
\tablecaption{Initial Dynamical Parameters\tablenotemark{a}}
\label{tab:dynamics} 
\tablewidth{0pt}
\tablehead{  
\colhead{parameter\tablenotemark{b}} & 
\colhead{orbit 1}   & 
\colhead{orbit 2}   & 
\colhead{} }
\startdata
Period  (days)&$ 1.72220593129863997E+00$ & $ 3.40173093143704293E+01$ & \cr 
$e\cos\omega$ &$-8.43853492103716768E-03$ & $-2.01600042188765399E-01$ & \cr 
$e\sin\omega$ &$ 8.13099536892672101E-03$ & $-2.40078982956226511E-01$ & \cr 
$i$  (rad) &$ 1.50838618564343707E+00$ & $ 1.61763510387093468E+00$ & \cr 
$\Omega$  (rad) &$-2.40688232452640587E-02$ & $-1.42436109017807011E-01$ & \cr 
$T_{\rm conj}$  (days)\tablenotemark{c} &$-3.42741046292364899E+01$ & $-1.08814170766013998E+01$ & \cr 
$a$  (AU) &$ 2.14262606526899012E-02$ & $ 2.45824516016162148E-01$ & \cr 
true anomaly  (deg) &$ 1.63544422724428642E+02$ & $ 3.45294078984964187E+02$ & \cr 
mean anomaly  (deg) &$ 1.63160805600938119E+02$ & $ 3.52662976217043195E+02$ & \cr 
mean longitude  (deg) &$ 2.98228699674208372E+02$ & $ 5.67112110508775572E+02$ & \cr 
\noalign{\vskip 2mm}\hline\noalign{\vskip 2mm}
\multicolumn{1}{c}{parameter\tablenotemark{d}} & \multicolumn{1}{c}{body 1} & 
\multicolumn{1}{c}{body 2} & \multicolumn{1}{c}{body 3} \cr
\noalign{\vskip 2mm}\hline\noalign{\vskip 2mm}
Mass ($M_{\odot}$) & $ 2.35185872607472007E-01$ & $ 2.07265802892892009E-01$ & $ 1.27021368536373869E+00$ \cr 
Radius ($R_{\odot}$) & $2.54533899427800026E-01$ & $2.31511559804099987E-01$ & $1.99826060683653917E+00$  \cr
Relative Flux & $4.3365469772E-04$ & $3.0362714803E-04$ & $9.9926271815425E-01$ \cr
$x$ (AU) & $ 9.64613800078151229E-02$ & $ 1.07134698565826331E-01$ & $-3.53418591401317775E-02$ \cr 
$y$ (AU) & $-1.73251011447649039E-02$ & $-1.87572885702252286E-02$ & $ 6.26852284626861107E-03$ \cr 
$z$ (AU) & $ 8.15747278537723303E-02$ & $ 6.27736588688604102E-02$ & $-2.53469606964325037E-02$ \cr 
$v_x$ (AU day$^{-1}$) & $-6.04984285813937109E-02$ & $ 6.87144201761893773E-03$ & $ 1.00803202784069835E-02$ \cr 
$v_y$ (AU day$^{-1}$) & $ 2.09269191790209747E-03$ & $ 2.83933194037935431E-03$ & $-8.50776528047805162E-04$ \cr 
$v_z$ (AU day$^{-1}$) & $ 1.83124994809432153E-02$ & $ 5.62022668666268821E-02$ & $-1.25613897255346038E-02$ \cr 
\enddata
\tablenotetext{a}{Reference time =       -35.00000,
integration step size = 0.01000 days}
\tablenotetext{b}{Jacobian instantaneous (Keplerian) elements}
\tablenotetext{c}{Times are relative to BJD 2,455,000.000}
\tablenotetext{d}{Barycentric Cartesian coordinates}
\end{deluxetable}

\section{Results}\label{sec:results}
\begin{figure}[ht!]
\centering
\includegraphics[width=\textwidth, trim=0 80 0 80, clip]{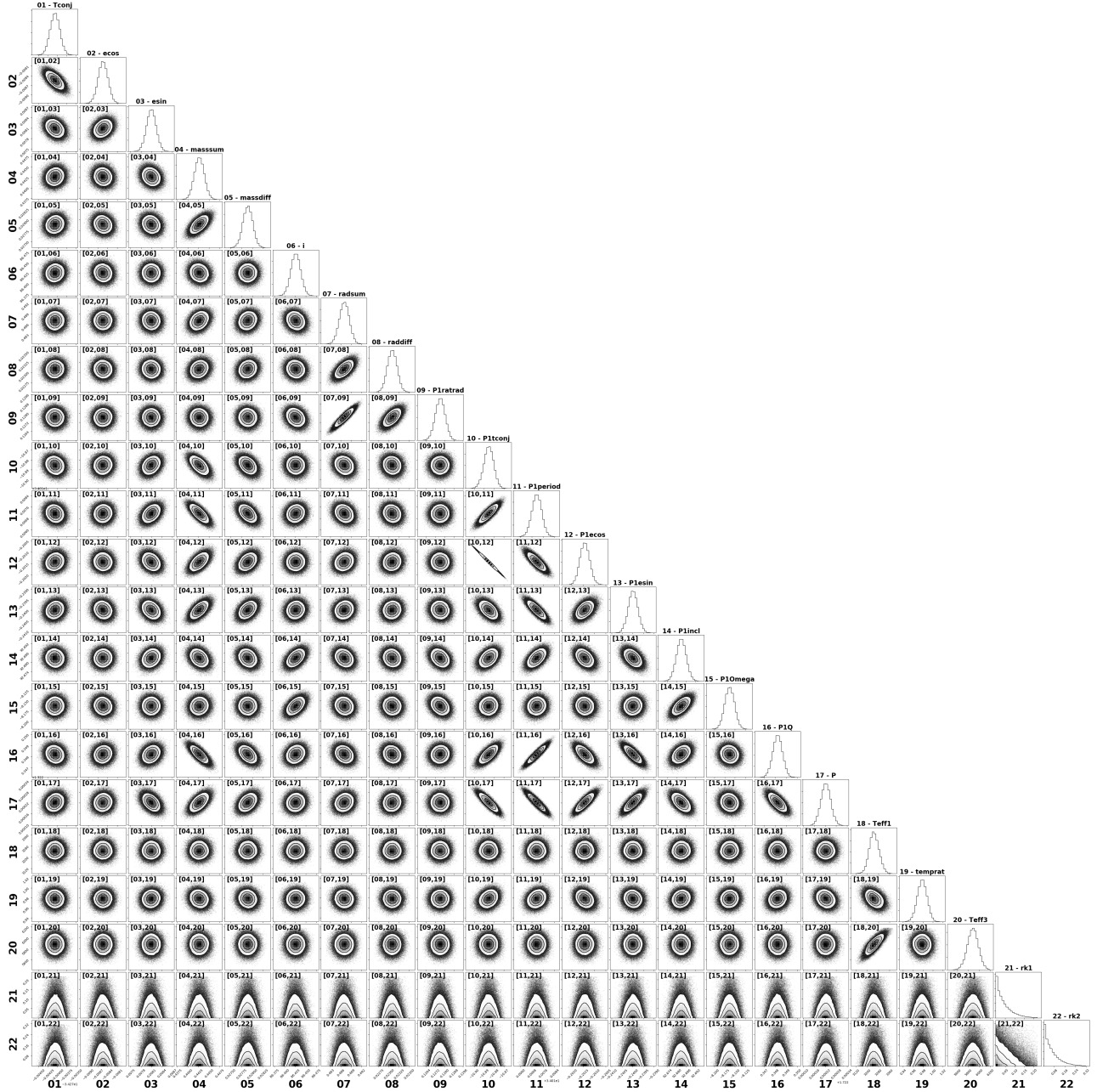}
\caption{Two parameter posterior plots for the fitting parameters from the final DEMCMC run \citep{2016JOSS....1...24F}.
The pair of tags for each plot matches those in Table \ref{tab:final-parms}.
Outlier models at the 4-$\sigma$ level were clipped from this plot,
amounting to 2286 of the 121520 samples. The contours contain the 11.8\%, 39.3\%, 67.5\%, and 86.5\%  of the population respectively.}
\label{fig:corner}
\end{figure}

The final values for the fitted parameters are given in Table \ref{tab:final-parms}, and other orbital and stellar parameters of interest that can be derived from the fitted parameters are given in Table \ref{tab:derived-parms}. The two-parameter posterior distributions showing the correlations 
of the fitting parameters can be seen in Figure \ref{fig:corner}.

\subsection{Stellar Masses, Radii, and Temperatures}\label{sec:mass-radius-temp}
The orbital configuration of KOI-126 allows extremely precise mass and radius determinations. We find stellar masses of 
M$_{A} = 1.2713 \pm 0.0047  ~M_{\odot} \ (0.37\%)$,
M$_{B} = 0.23529 \pm 0.00062 ~M_{\odot} \ (0.26\%)$, and
M$_{C} = 0.20739 \pm 0.00055 ~M_{\odot} \ (0.27\%)$.
The stellar radii are 
R$_{A} = 1.9984 \pm 0.0027 ~R_{\odot} \ (0.14\%)$,
R$_{B} = 0.25504 \pm 0.00076 ~R_{\odot} \ (0.30\%)$, and
R$_{C} = 0.23196 \pm 0.00069 ~R_{\odot} \ (0.30\%)$.
\citet{2011Sci...331..562C} made the optimistic prediction that when all the \ik~data are included, the relative uncertainties in mass and radius for stars B and C will be determined to better than 0.1\%. While our results do not reach that level of precision, these are still among the most precisely measured values for late type stars. 

Comparing these values to those found by \cite{2011Sci...331..562C}
(M$_{A} = 1.347   \pm 0.032 ~M_{\odot}$,
M$_{B} = 0.2413 \pm 0.0030 ~M_{\odot} $,
M$_{C} = 0.2127 \pm 0.0026 ~M_{\odot} $;
and
R$_{A} = 2.0254 \pm 0.0098 ~R_{\odot}$,
R$_{B} = 0.2543 \pm 0.0014 ~R_{\odot}$,
R$_{C} = 0.2318 \pm 0.0013 ~R_{\odot}$)
we find differences at the $\sim$2-$\sigma$ level in the masses of the stars,
and for the radius of KOI-126 A, though interestingly, the radii of KOI-126 B and C are in close agreement. 
These mild disagreements are not surprising given that we are using updated radial velocities, and that we also include four seasonal background-light contamination parameters for the \ik~light curves. We also explicitly include classical apsidal motion, which \cite{2011Sci...331..562C} did not include in their published solution, and we employ an MCMC methodology to determine the uncertainties on the parameters, not a Levenberg-Marquardt minimization covariance matrix. Finally, \cite{2011Sci...331..562C} had access to 247 d of \ik~observations, which covers less than one apsidal precession cycle of the B+C inner binary, 
while our \ik\ and MLO eclipse observations span slightly over 2 cycles.

For the effective temperature of KOI-126 A we adopt the value from the spectroscopic analysis by \citet{2011Sci...331..562C}: $\rm{T_{eff}} = 5875 \pm 100$ K.  
Stars B and C are much too faint to get a direct spectroscopic temperature measurement, but the depths of the eclipses do allow a constraint on their luminosities and with their radii, the temperatures.
For KOI-126~A the luminosity is calculated using the posterior values for the radius and the effective temperature. The effective temperatures of KOI-126 B and C are constrained by the ratio of the stellar flux, which are determined by fitting eclipses. 

\begin{figure}[ht!]
\centering
\includegraphics[width=\textwidth, trim=0 40 0 40, clip]{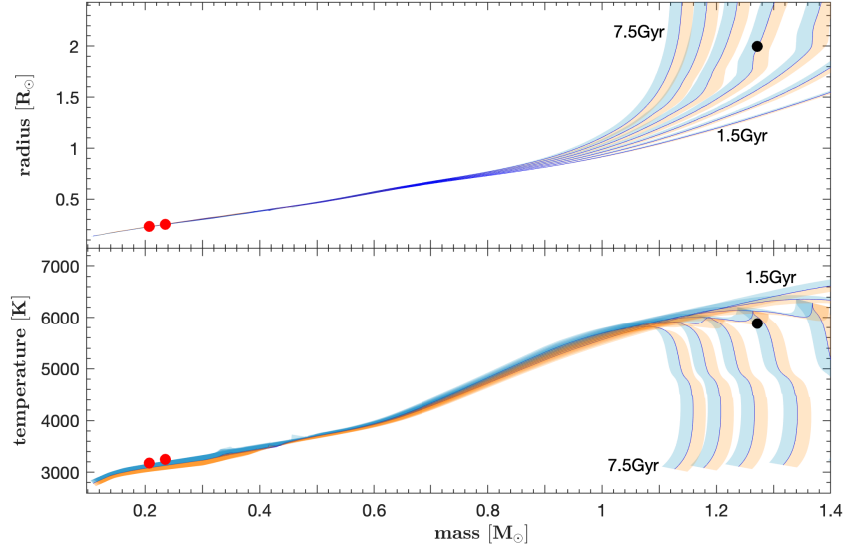}
\caption{The measured masses, radii, and temperatures of KOI-126 are shown along with the MIST isochrones spanning 1.5 to 7.5 Gyrs in steps of 1 Gyr. 
A metalicity of ${\rm [Fe/H]=0.15}$ was used when using MIST to generate all of the isochones. The shaded regions around each isochrone curve shows the effect of the $\pm 0.08$ dex uncertainty in the metallicity: the cream color corresponds to 
${\rm [Fe/H]=0.23}$ and the light blue to ${\rm [Fe/H]=0.07}$. 
Star A falls on the 4.5 Gyr isochrone for both the R vs.\ M and the ${\rm{T_{eff}}}$ vs.\ M planes. The uncertainties are plotted but are are smaller than the size of the symbols.
}
\label{fig:isochrone}
\end{figure}

\begin{figure}[ht!]
\centering
\includegraphics[width=\textwidth, trim=0 80 0 80, clip]{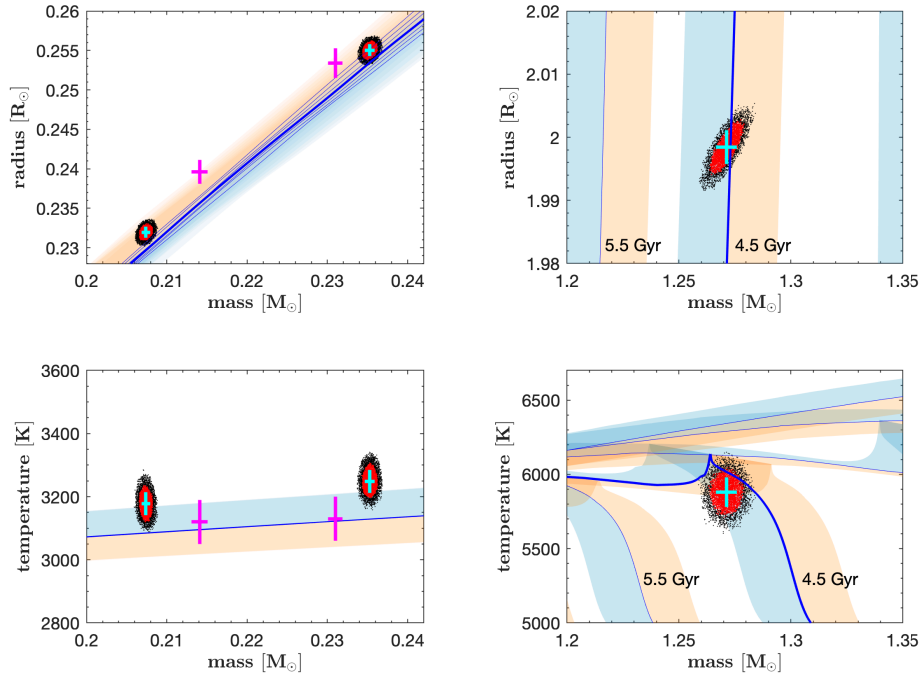}
\caption{Similar to Figure \ref{fig:isochrone} but zoomed in on star A (righthand panels) and stars B and C (lefthand panels). The cloud of points illustrate the posterior distribution from the MCMC modeling, with the red color denoting the 68.3\% probability range (1-$\sigma$) and the black denoting the 95.5\% (2-$\sigma$) range. The standard uncertainties are shown as cyan-colored error bars. For comparison, the pair of magenta crosses in the left panels are for the well-characterized eclipsing binary CM Dra.
}
\label{fig:isochrone_zoom}
\end{figure}

We compared our measured masses, radii, and temperatures with theoretical stellar evolutionary models from the Mesa Isochrones and Stellar Tracks - MIST
\citep{2016ApJS..222....8D,2016ApJ...823..102C,2011ApJS..192....3P,2013ApJS..208....4P,2015ApJS..220...15P,2018ApJS..234...34P}.
The specific models were retrieved from the MIST web interpolator for the 
metallicity measured for star A by \citet{2011Sci...331..562C}: ${\rm [Fe/H]=0.15}$. 
As shown in Figures \ref{fig:isochrone} and \ref{fig:isochrone_zoom}, in both the mass-radius and mass-temperature planes, star A is located 
on the 4.5 Gyr isochrone curve. Including the uncertainty in metallicity ($\pm 0.08$ dex) results in an age of $4.5 \pm 0.3$ Gyrs.
This is consistent with the $4 \pm 1$ Gyr age estimated by \citet{2011Sci...331..562C}, and the 3-5 Gyr estimate of \citet{2012MNRAS.422.2255S}. In Figure \ref{fig:isochrone_zoom}, in addition to the 1-$\sigma$ error bars, we show a small (5\%) random subset of the DEMCMC posterior sample from the photodynamical modeling for the masses, radii, and temperatures. The posterior sample reveals the correlations between parameters that standard error bars do not.

Stars B and C, being so low in mass, do not provide any constraint on the age, but they are nonetheless very interesting. Given their masses, they are both expected to have fully-convective interiors
and they do
exhibit the well-known inflated radius discrepancy between evolutionary models and observations (e.g.\ see \cite{2006ApJ...651.1155B,2019MNRAS.482.5379G,2010AARv..18...67T,2013ApJ...776...87S}). 
Our result is in slight disaccord with the findings of \citet{2012MNRAS.422.2255S} who find that for a metallicity between solar and +0.3 the 
{\textsc{phoenix}} 
NextGen  1D model atmospheres can well-match the observations -- though they are using the 
\citet{2011Sci...331..562C} stellar values for KOI-126.
This discrepancy is reminiscent of
the very well-determined binary star system CM Dra, which has stars of similar mass to KOI-126 B and C yet their radii are significantly larger and cannot be reconciled with standard theoretical models using the inferred low metallicity for the stars \citep{2009ApJ...691.1400M}. This disagreement is even more notable when taking into consideration that the stars in CM~Dra should be smaller in radius than KOI-126 B and C because of the much lower metallicity in CM~Dra. The reason for this discrepancy is not understood -- see \cite{2011ApJ...740L..25F} for a discussion of possible causes, though we note that it is unlikely stars B and C in KOI-126 are in a period of low magnetic activity as postulated in \cite{2011ApJ...740L..25F} as a possible cause for their smaller radii since, unlike the \citet{2011Sci...331..562C} study, we are using data that span nearly a decade. 

\subsection{Orbital Precession}\label{sec:precession}
\begin{figure}[ht!]
\centering
\plotone{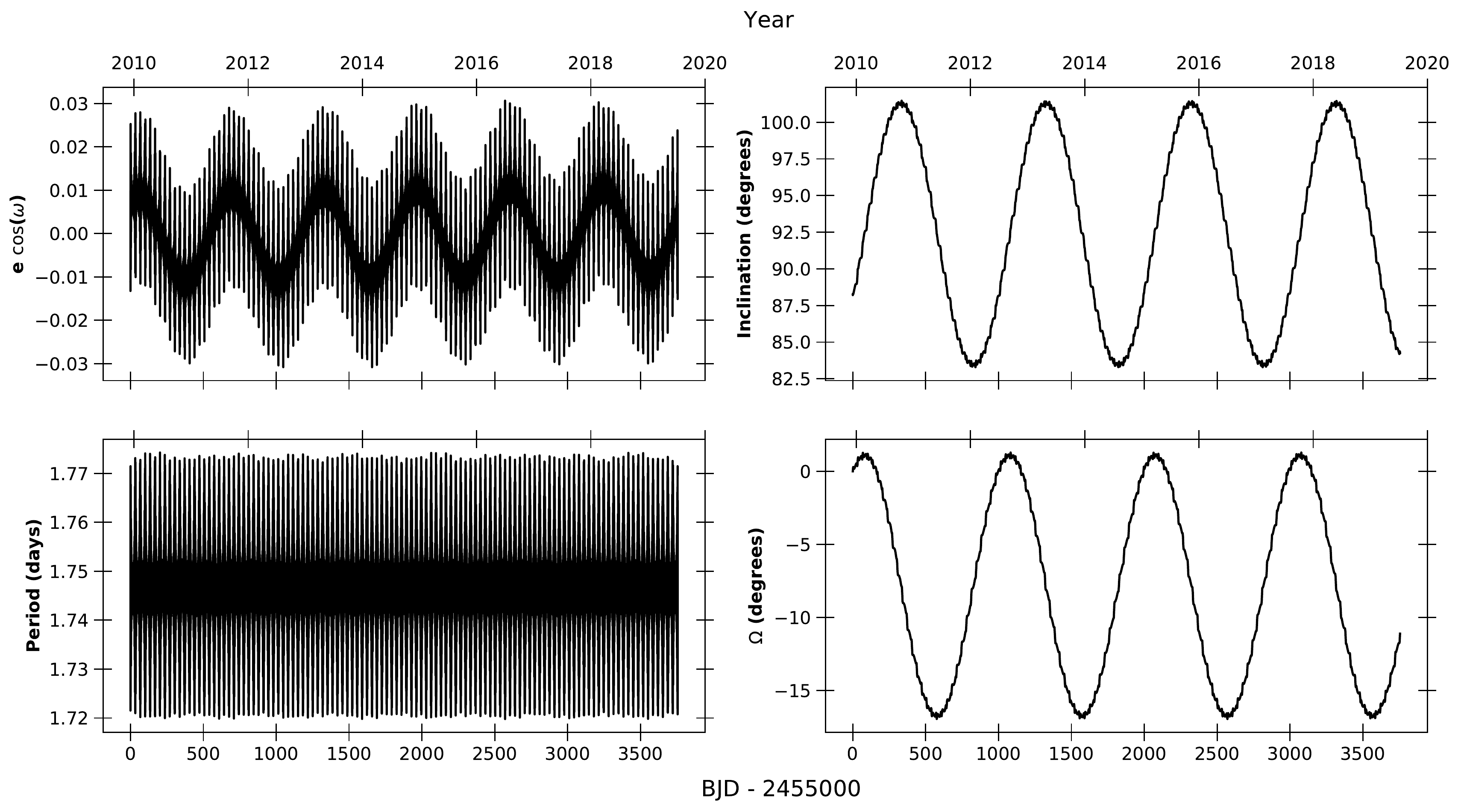}
\caption{The osculating orbital parameters of the inner orbit (KOI-126 B and C) over the range of our data set. Periodicities consistent with the mean period of the outer and inner orbit are present in each. An apsidal period of 1.741 years can be seen in the $e\cos\omega$ plot. 
The inclination and $\Omega$ plots have a periodicity of 2.73 years. }
\label{fig:inner}

\plotone{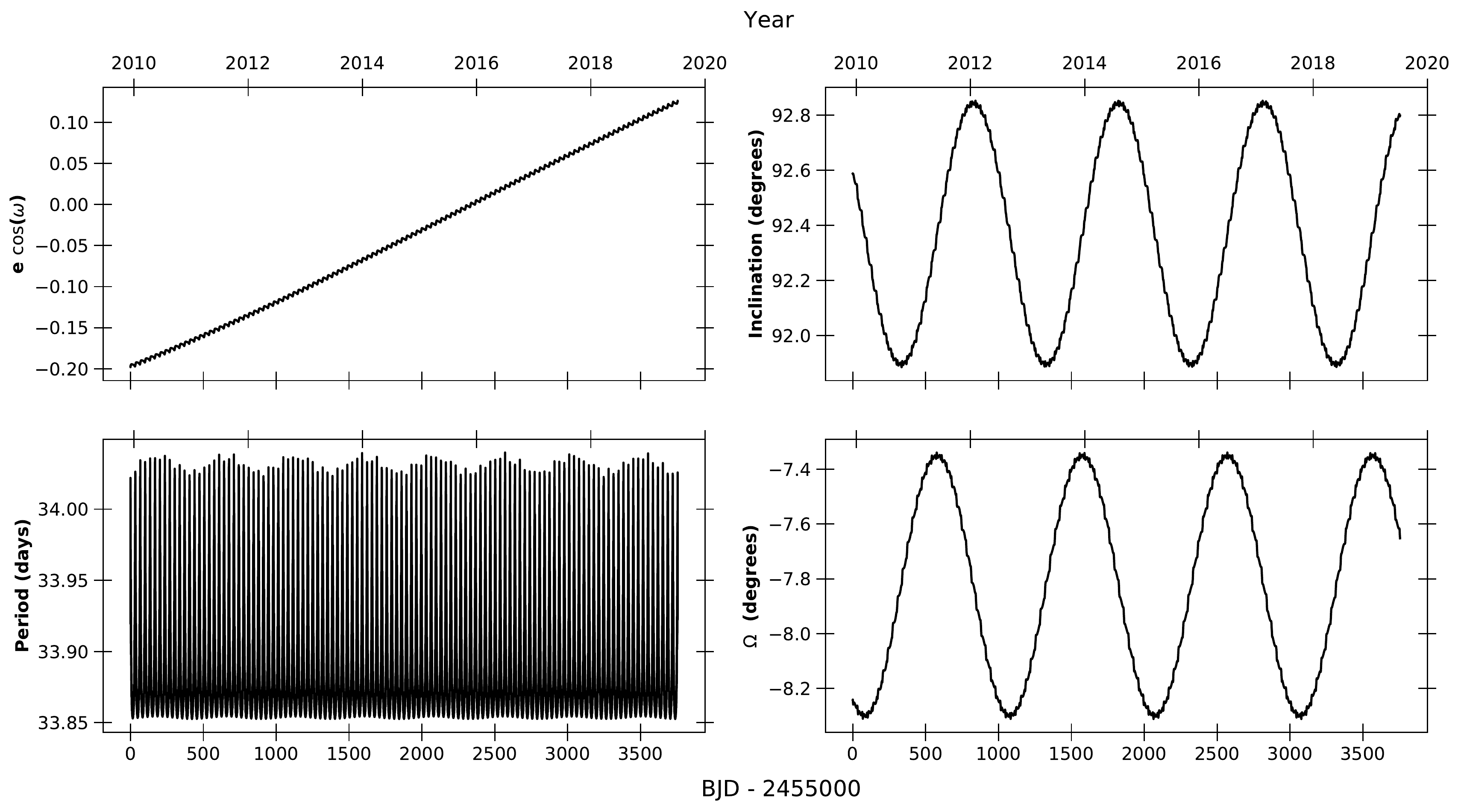}
\caption{The osculating orbital parameters of the outer orbit (KOI-126 A around the barycenter). Periodicities consistent with the mean period of the outer and inner orbit are present in each.
An apsidal period of 57.9 years can be seen in the $e\cos\omega$ plot. 
The inclination and $\Omega$ plots have a periodicity of 2.73 years. }
\label{fig:outer}
\end{figure}

Due to the 3-body orbital dynamics, the instantaneous Keplerian elements change rapidly, and the values reported in these tables are valid only for the epoch T$_{ref}$ = 2454965 BJD. 
To estimate periodicities in the orbital parameters, a model using the best fitting parameters was generated for a span of 50,000 days ($\sim$137 years)
with time steps of 0.01 days and parameters reported every 0.1 days. 
Figures \ref{fig:inner} and \ref{fig:outer} show how the $e\cos\omega$, inclination, orbital period, and nodal angle ($\Omega$) of the inner and outer orbit evolve over the time span of our observations. In particular, the value of $e\cos\omega$ is closely related to the precession of the argument of periastron, which corresponds to our sought-after apsidal precession rate. There are multiple periodicities present in the orbital elements shown in 
Figures \ref{fig:inner} and \ref{fig:outer}, and using the full 50,000-day
simulation a Lomb-Scargle periodogram was used to determine their periods \citep{1992nrfa.book.....P}.
As expected, periodicities at the inner and outer orbital periods are present in each orbital element. 
More interestingly, from the $e\cos\omega$ curves the apsidal precession rate for the B+C inner orbit is a remarkably fast 
636.0 days (1.741 years). 
This is in good agreement with  \citet{2011Sci...331..562C} who found a period of $\sim$650 days.
The outer orbit's apsidal precession rate is much longer at 
21,850 days (59.8 years). 
The inner orbit also exhibits a small residual periodicity that matches the outer orbit's precession rate. 
The inclination and nodal angle of the inner and outer orbit are related, and have identical precession periods of 
996 days (2.73 years). 
This value agrees well with the $\sim$950 day period found by \citet{2011Sci...331..562C}.

\begin{figure}[ht!]
\plotone{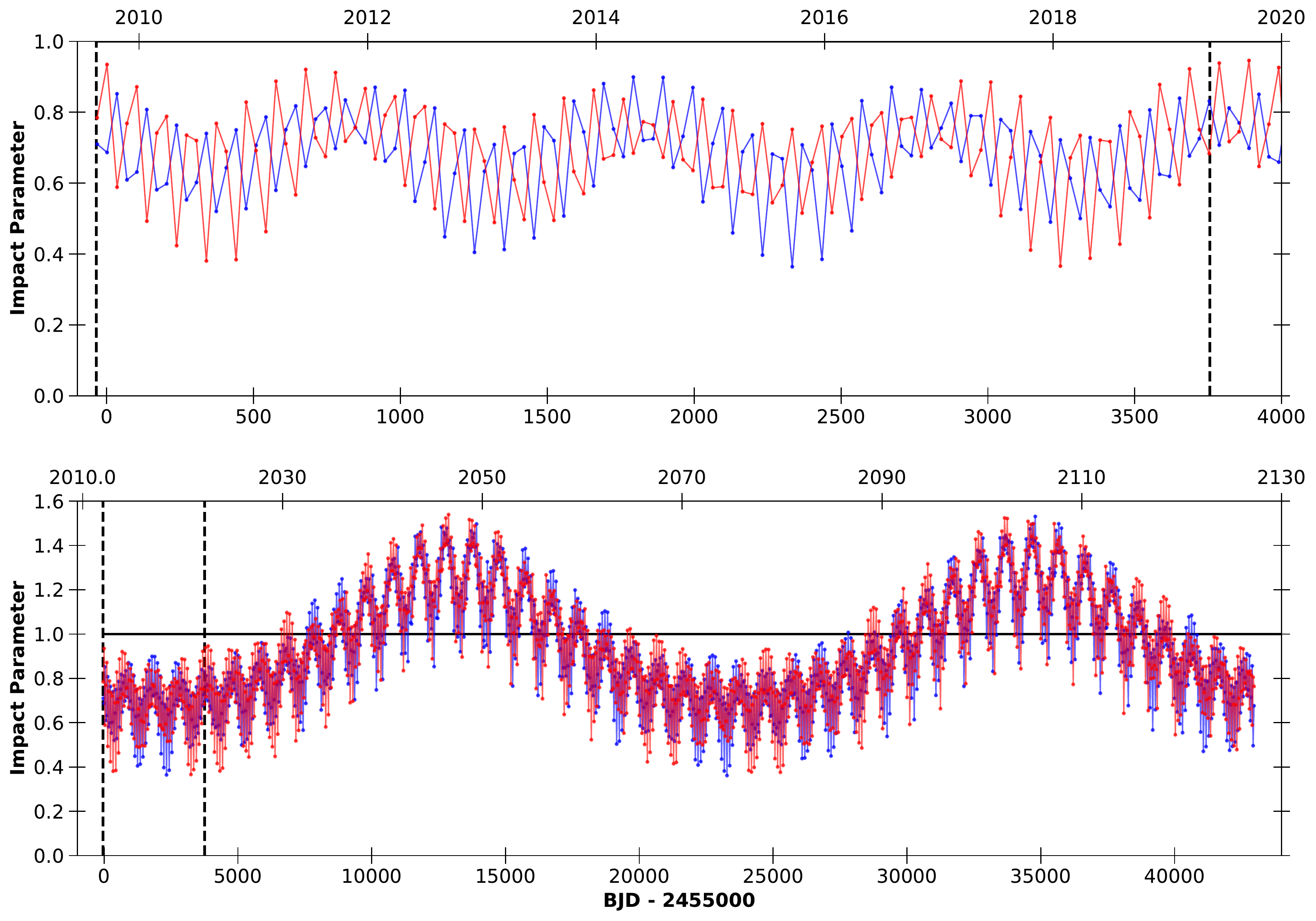}
\caption{The absolute value of the impact parameters of KOI-126 B (blue) and KOI-126 C (red) at times of minimum on-sky separation. The bottom plot shows the changing impact parameter over 2 cycles of 43,000 days with a solid line indicating an impact parameter of 1. In both plots the vertical dashed lines indicate the range of the data used in this study.
}
\label{fig:impact}
\end{figure}

As the inclination and argument of periastron for the inner and outer orbits evolve, the complex eclipse profiles will change, mostly due to the changing impact parameter and relative velocities of the bodies. Figure \ref{fig:impact} shows the change in the impact parameter over the range of our observations in the upper panel and out to 43,000 days (=$\sim$2 precession cycles of the outer orbit) in the lower panel. At the current time (including all the \ik~observations) the impact parameter is always less than 1.0, meaning an eclipse will occur at every conjunction.
However, as the orbits precess, KOI-126 B and C will sometimes be orientated such that the conjunctions do not result in eclipses. Over a precession cycle of the outer orbit (21505 days) around 68\% of the conjunction will result in an eclipse. During the 14 years between BJD 2,465,000 and 2,470,113 (2036 Nov to 2050 Nov), very few of the conjunction will produce eclipses, with a majority of those occasional eclipses being shallow grazing eclipses. 

\subsection{The Apsidal Motion Constants}\label{sec:k2}
In their pioneering work, \citet{2011Sci...331..562C} were able to measure the apsidal motion constants $k_{2}$ of the M-stars KOI-126 B and C to be less than 0.6. While this weak constraint is in itself is not particularly valuable, they further stated that the determination of $k_{2}$ to 1\% precision should be achievable when the system was analysed using the full \ik~data. This largely motivated our work on KOI-126 -- being able to constrain the density distribution inside a low-mass star would be of considerable importance.

Because the stars are of identical age and composition and they have very similar mass and radius, their apsidal motion constants are expected to be very nearly the same. \citet{2011ApJ...740L..25F} estimate that the values for $k_{2,B}$ and  $k_{2,C}$ should be 0.149 and 0.151 respectively, using the stellar parameters of \citet{2011Sci...331..562C}. In our photodynamical modeling we let $k_{2}$ be a free parameter for each M-star, as this gave us more freedom to identify correlations, though we are aware that there is a very strong degeneracy and only the sum (or average) of the apsidal motion constants can be constrained by the orbital precession rate.

\begin{figure}[ht!]
\centering
\includegraphics[scale=0.4]{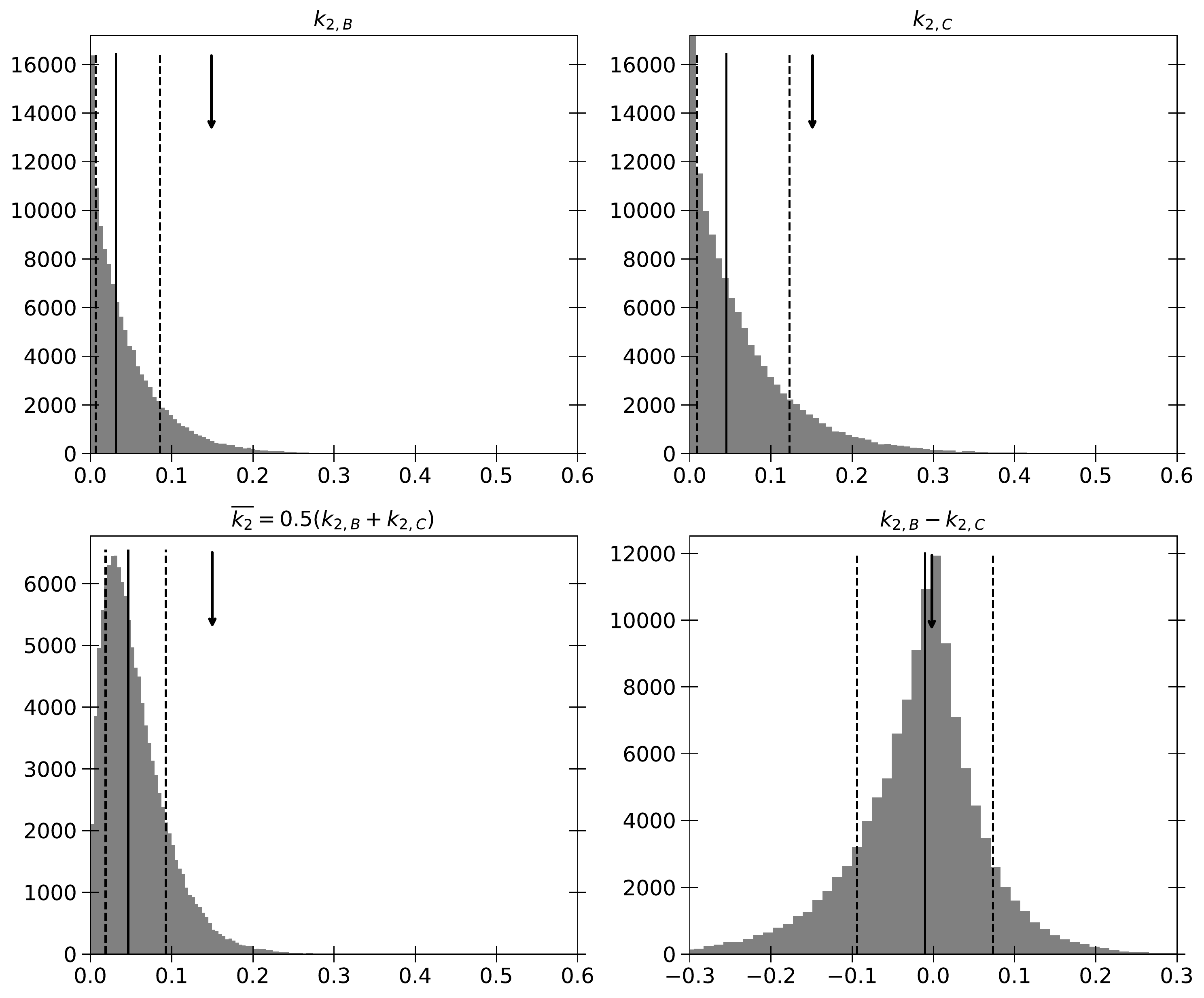}
\caption{Posterior distributions of the individual and unweighted mean apsidal motion constants
for KOI-126 B and C ($k_{2,B}$, $k_{2,C}$, and $\overline{k_{2}}$) are shown. For completeness the difference between $k_{2,B}$ and $k_{2,C}$ is also plotted. In each plot the solid black line indicates the median value,the dashed lines are the 1 $\sigma$ uncertainties, and the arrows indicate the theoretical values.}
 \label{fig:rk_hist}
\end{figure}

As stated in the previous section, the orbits do precess on a rapid timescale, giving us hope that the apsidal motion constants can be readily determined. Including $k_2$ for each star was relatively straightforward in our modeling, and the individual posterior distributions for these parameters, as well as the derived posterior distribution of their mean value are shown in Figure \ref{fig:rk_hist}.
The result is surprising -- the individual posterior distributions (top panels of Figure \ref{fig:rk_hist}) have a strong peak at 0.0. 
The medians of these one-sided distributions 
$k_{2,B} = 0.031^{+0.055}_{-0.025}$, and 
$k_{2,C} = 0.045^{+0.078}_{-0.036}$
are slightly
closer to the expected theoretical values ($\sim$0.15), but still disagree at the $\sim$2-$\sigma$ level
($\sigma$ is defined here such that $\pm \sigma$ encompasses 68.3\% of the distribution).
Perhaps more telling, the average of the sum of the apsidal motion constants 
$\overline{k_{2}} = 0.046^{+0.046}_{-0.028}$
is not peaked at zero and its median is closer to the expected value.
This combination of $k_2$ values is our final solution, as it is the average that is constrained by the observation of the orbital precession of the apsides, but it remains roughly 2-$\sigma$ low compared to the the theoretical expectation of  \citet{2011ApJ...740L..25F}.
The relative uncertainty in the average $\overline{k_{2}}$ is 
only 61-100\%, and it is a significant improvement over the
upper limit of $\overline{k_{2}} < 0.6$ by \citet{2011Sci...331..562C}, but it is far from the predicted 1\% precision that was hoped for.

Since KOI-126 B and C are within the fully convective regime, the predicted $k_{2}$ of 0.15, or the equivalent polytropic index of $\sim$1.5, gives a ratio of core density to mean density of $\sim$3.3. Our inferred $\overline{k_{2}}$ value of $\sim$0.046 implies a polytropic index 
between 3 and 3.5, and a ratio of core density to mean density in the range of 54 to 152; these would be much more centrally dense stars \citep{2009ApJ...698..715S,1955MNRAS.115..101B}. 
However, these $k_2$ values and their implications should not necessarily be taken at face value
because of their relatively large uncertainties and potential systematic biases.
In the next section we explore possible reasons for the low $k_2$ values and perform tests on their reliability. 

\begin{deluxetable}{clccccc}
\tabletypesize{\scriptsize}
\tablecaption{KOI-126 Fitting Parameters\label{tab:final-parms}}
\tablewidth{0pt}
\tablehead{
\colhead{tag} &
\colhead{ Parameter } &
\colhead{Unit} &
\colhead{Best Fit} &
\colhead{Median} &
\colhead{+1 $\sigma$} &
\colhead{-1 $\sigma$} 
}
\startdata
01 & T$_{conj, 1}$ &   BJD - 2455000   &   -34.27411   &  -34.27407  &       0.00013     &   0.00013 \\
02 & $e_{1}\cos\omega_{1}$ &       &   -0.00844    & -0.00855     &    0.00015   &     0.00015  \\
03 & $e_{1}\sin\omega_{1}$ &        &  0.00813   & 0.00809    &     0.00014    &    0.00014   \\
04 & M$_{B}$ + M$_{C}$  &     M$_{\odot}$   &  0.4425 & 0.4427  &      0.0012  &      0.0012 \\
05 & M$_{B}$ - M$_{C}$  &      M$_{\odot}$   & 0.02792 & 0.02790  &       0.00011   &     0.00011  \\
06 & i$_{1}$ &   degrees    &   86.424  & 86.426    &     0.012    &    0.012 \\
07 & R$_{B}$ + R$_{C}$  &    R$_{\odot}$   & 0.4861   & 0.4870   &      0.0014    &    0.0014  \\
08 & R$_{B}$ - R$_{C}$  &  R$_{\odot}$   & 0.0230 & 0.02308  &       0.00018   &     0.00012  \\
09 & R$_{A}$/R$_{C}$  &     & 0.12738  & 0.12762     &    0.00042    &    0.00043 \\
10 & T$_{conj, 2}$ &   BJD - 2455000    & -10.8814 & -10.8854   &      0.0049   &     0.0050  \\
11 & P$_{2}$ &    days   & 34.01731  & 34.01714  &       0.00034   &     0.00034  \\
12 & $e_{2}\cos\omega_{2}$ &       & -0.20160   & -0.20143   &      0.00022  &      0.00021 \\
13 & $e_{2}\sin\omega_{2}$ &       & -0.24008   & -0.23999    &     0.00023   &     0.00023  \\
14 & i$_{2}$ &      degrees  & 92.6837  & 92.6828    &     0.0031   &    0.0031  \\
15 & $\Omega_{2}$&  degrees     & -8.161 &  -8.163    &     0.018     &   0.018  \\
16 & (M$_{B}$+M$_{C}$)/M$_{A}$ &   & 0.34833 & 0.34822    &     0.00043   &     0.00043  \\
17 & P$_{1}$ &    days   & 1.722206 &   1.722223    &     0.000027    &    0.000027  \\
18 & Teff$_{B}$ &     K   & 3235  &  3248     &   36   &    33     \\
19 & T$_{eff,C}$/T$_{eff,B}$    &   &  0.9796  & 0.9782    &     0.0097   &     0.0097 \\
20 & T$_{eff,C}$  &     K   &  5840 & 5880     &   99   &    100   \\
21 & $k_{2,B}$ &          &  0.023 &  0.031     &    0.054    &    0.025   \\
22 & $k_{2,C}$ &          & 0.001   &  0.045     &    0.078    & 0.036   \\
23* & \ik~Q1 &     &  0.369    &  0.366     &    0.011     &   0.011      \\
24* & \ik~Q2 &     &  0.293  &   0.296     &    0.015    &    0.014      \\
25* & MLO R-band Q1 &    & 0.2003   &     0.2033     &    0.0053    &    0.0025       \\
26* & MLO R-band Q2 &   & 0.399 &      0.389    &     0.073    &    0.091      \\
27$^\dagger$ & S0 contamination &  & 0.0241 &  0.0274     &    0.0059   &     0.0061     \\
28$^\dagger$ & S1 contamination &  &  0.0363 & 0.0397    &     0.0058    &    0.0060     \\
29$^\dagger$ & S2 contamination &  &  0.0480 &   0.0514      &   0.0058    &    0.0059     \\
30$^\dagger$ & S3 contamination &  &  0.0243 &  0.0278      &   0.0058    &    0.0060      \\
\enddata       
\tablenotetext{}{
The subscripts $_{1}$ and $_{2}$ refer to the inner orbit (B + C) and the outer orbit (B + C around A) respectively.\\
* Q1 and Q2 are the \cite{2013MNRAS.435.2152K} quadratic limb darkening law coefficients for KOI-126 A in the \ik~and R bandpasses. These parameters are not plotted in Figure \ref{fig:corner}.\\
$^\dagger$ S0, S1, S2, and S3 contamination are the seasonal flux contamination parameters for the \ik~data. These parameters are not plotted in Figure \ref{fig:corner}.}
\end{deluxetable}

\begin{deluxetable}{lccccc}
\tabletypesize{\scriptsize}
\tablecaption{KOI-126 Derived Parameters \label{tab:derived-parms}}
\tablewidth{0pt}
\tablehead{
\colhead{ Parameter } &
\colhead{Unit} &
\colhead{Best Fit} &
\colhead{Median} &
\colhead{+1 $\sigma$} &
\colhead{-1 $\sigma$} 
}
\startdata
inner orbit (B+C) &    &      &      &       &    \\
M$_{B}$ & M$_{\odot}$ &   0.23519    &  0.23529    &   0.00062     &   0.00062  \\
M$_{C}$ & M$_{\odot}$ &  0.20727    &   0.20739     &    0.00055  &      0.00055 \\
R$_{B}$ & R$_{\odot}$ & 0.25453     &   0.25500    &     0.00075  &      0.00076  \\
R$_{C}$ & R$_{\odot}$ &  0.23151     &   0.23196     &     0.00067  &      0.00069  \\
P$_{\text{1}}$  & days &  1.722206   &   1.722223    &  0.000027    &    0.000027  \\
a$_{\text{1}}$  & AU &  0.021426     &   0.021557     &0.000019     &   0.000019  \\
e$_{\text{1}}$   &   &   0.01172  &    0.01177     &    0.00012   &     0.00012   \\
$\omega_{\text{1}}$  & degrees & 136.06 &   136.56  &     0.80   &     0.80 \\
i$_{\text{1}}$  & degrees  &    86.424  &  86.426    &     0.012    &    0.012  \\
log $g_{B}$ & cgs &  4.998 &   4.997      &  0.0025    &    0.0025  \\
log $g_{C}$ & cgs & 5.0255 &   5.0240     &     0.0025   &     0.0025  \\
$\rho_{B}$ &  g/cc &  20.11&   20.00     &   0.17   &     0.17 \\
$\rho_{C}$ &  g/cc &   23.55  &     23.43  &       0.20  &      0.19\\
T$_{eff, B}$ &  K &  3235  &    3248     &   36   &    33 \\
T$_{eff, C}$ &  K & 3169     &    3177   &    37   &    35 \\
L$_{B}$ &  L$_{\odot}$ &  0.00639  &   0.00653 &     0.00030     &   0.00026 \\
L$_{C}$ &  L$_{\odot}$ &  0.00487 &  0.00495  &     0.00023   &     0.00022  \\
$\overline{k_{2}}$ &  &  0.025     &   0.046   &      0.046     &   0.028 \\ \hline 
outer orbit (A+(B+C))&        &    &    &   &    \\
M$_{A}$ & M$_{\odot}$ & 1.2702      &   1.2713   &   0.0047   &     0.0047  \\
R$_{A}$ & R$_{\odot}$ &   1.9983    &   1.9984    &     0.0027    &    0.0026  \\
P$_{\text{2}}$  & days &  34.01731     &    34.01714   &   0.00034    &    0.00034   \\
a$_{\text{2}}$   &  AU  &  0.245828     &      0.245888  &      0.000278    &    0.00028    \\
e$_{\text{2}}$   &    &  0.31350    &   0.31332   &  0.00027   &     0.00027  \\
$\omega_{\text{2}}$  & degrees  &  229.979    &    229.993  &     0.030    &    0.029   \\
i$_{\text{2}}$  & degrees &  92.6837   &  92.6828   &       0.0031     &   0.0031   \\
log $g_{A}$ & cgs &   3.9406   &   3.9409     &    0.0011     &   0.001   \\
$\rho_{A}$ & g/cc &   0.22443 &     0.22457     &    0.00060      &  0.00060 \\
T$_{eff, A}$ &  K &   5840  &    5880 & 99   &   100 \\
L$_{A}$ &  L$_{\odot}$ &  4.19 &     4.30  &  0.29  &    0.29  \\
\enddata
\tablenotetext{}{The reported Keplerian parameters are the instantaneous
(osculating) values at the reference epoch T$_{\text{ref}}$=2,454,965 BJD. 
Figures \ref{fig:inner} and \ref{fig:outer} show how these elements evolve over time.}
\end{deluxetable}

\section{Discussion}\label{sec:discussion}
\subsection{Simulated Data Tests}\label{sec:simulated}
Given the surprising results of our fitting for the apsidal motion constants for KOI-126 B and C, we decided to explore and test the inherent limitation of our photodynamical modeling.  Two simulated data sets were generated using the best fitting model:
One had the $k_{2}$ values set to 0, the other had them set to their expected
values of 0.149 and 0.151 \citep{2011ApJ...740L..25F}. These two forward models were sampled exactly like the \ik, MLO, and RV datasets so each datum in the simulation corresponded to an observed datum,
though the noisy occultation segments were omitted for convenience, since they add many data points but provide only a weak constraint on the solution.
Artificial observational noise was then added to the simulated data by offsetting it by a random number drawn from a Gaussian distribution with a mean of zero and with a standard deviation matching the median \ik~or RV uncertainty. Two more simulated sets were generated through the same process, one for $k_2=(0,0)$ and one for $k_2 =(0.149,0.151)$, except now the simulated observational noise had a standard deviation 100 times smaller. In total, 4 simulated data sets were created.

These 4 datasets represent idealized scenarios in which we know the answer beforehand and the noise is perfectly Gaussian. These data were each fit in an identical way as the real data. Specifically, the DEMCMC chains were initialized with the same random number seed, the same starting population of points, and the same starting best fit model. Each was allowed to ``converge'' (i.e.\ reach a steady state) and the posterior analysis was carried out after the burn-in period. 

For the cases where the apsidal motion constants were set to zero we find that the individual $k_{2}$ posterior distributions are flat, with no preferred trend towards 0. When the individual $k_2$ are averaged together, the distribution peaks halfway between the upper and lower bounds of the allowed range in the fitting, consistent with drawing numbers uniformly between the upper and lower limits. This is indicative of completely unconstrained values of $k_{2}$. While these two tests did not recover the individual input values of $k_2$, they do show that our fitting method does not have a bias towards a value of zero.

For the case of $k_2$ set to their theoretically expected values and with noise consistent with the real observations, the individual posterior distributions exhibit a trend towards 0, similar to what is seen in our actual data. When the average $k_{2}$ distribution is examined, we find a peak at $\sim$0.2, somewhat larger than the input value of 0.15, but the distribution contains 0.15 within one-sigma.

A different result was seen in the very low-noise case. The individual $k_{2}$ values diverged, with one trending towards 0 and the other towards $\sim$0.27. However, the average $\overline{k_{2}}$ was tightly peaked around 0.135, close to the expected value of 0.15, but formally more than 2-sigma away because the distribution is narrow. The narrow peak is likely a result of the very small amount of noise added to generate the simulated data.
   
These tests suggest the following: (i) Somewhat predictably, the individual $k_2$ values are poorly determined and should not be considered in isolation separately; (ii) The input average $\overline{k_{2}}$ value is approximately recovered when not set to zero and the artificial data have a noise level that matches the observations. From these tests we conclude that the average $\overline{k_{2}}$ can be somewhat constrained by the KOI-126 observations, but only weakly so, and we find no evidence that the photodynamical fitting/modeling methodology biases the inferred $k_2$ values low.

\subsection{Exploring Possible Causes of a Low $k_2$ Value}\label{sec:shakura}
Modeling our simulated datasets has given us some confidence that the low $\overline{k_{2}}$ value we have determined is plausible, but it is also possible that other factors not included in our simulation or modeling code can lead to a spuriously low value. In the next subsections, we examine some of our assumptions.

\subsubsection{Spin and Orbit Axes Misalignment}
In our photodynamical modeling we assume 
that the B and C stellar rotation axes are aligned perpendicular to the orbital plane. 
This common assumption made in apsidal motion studies usually has a negligible effect on the apsidal precession rate. However, this is not always the case. DI Her is the most famous example of a binary system undergoing apsidal precession at a significantly slower rate than expected \citep{1984ApJ...287L..77M,1985AJ.....90.1519G}, which, for a time, led researchers to question the
validity of the precession rate due to general relativity. However, the slow precession rate was explained by the rotation axes of both stars being nearly in the orbital plane \citep{2009Natur.461..373A}, as measured using the Rossiter-McLaughlin effect \citep{1924ApJ....60...15R}.  Thus we explore the possibility that the lower-than expected apsidal motion constants (leading to a slower-than expected precession rate) in KOI-126 could be due to spin-orbit misalignment.

Because stars are not point masses, or even spherical, the orbital motion of the stars in a close binary star system is not strictly Keplerian. Stellar rotation causes stars to become oblate and the tidal force by the companion star creates a quadrupole moment that leads to precession of the star's orbit. In addition to this classical apsidal motion, there is the well-known general relativistic precession. \citet{1985SvAL...11..224S} gives an equation describing the precession that explicitly includes the stellar spin axis alignment:

\begin{equation}\label{eq:shakura}
\begin{aligned}
    \dot{\omega} & \ = \ \dot{\omega}_{gr} \ + \ \overline{\omega}15 g(e) \left [  k_{1} r^{5}_{1} \frac{M_{1}}{M_{2}} + k_{2} r^{5}_{2} \frac{M_{2}}{M_{1}} \right] \
   - \ \frac{\overline{\omega}}{\sin^{2}{i} (1-e^{2})^{2}} \times \\
   & \bigg\{ \ k_{1} r^{5}_{1} \left(\frac{\omega_{1}}{\overline{\omega}}\right)^{2} \left(1 + \frac{M_{2}}{M_{1}}\right) 
     \left [ \cos{\alpha_{1}}(\cos{\alpha_{1}} - \cos{\beta_{1}}\cos{i}) + \frac{1}{2}\sin^{2}{i} (1 - 5\cos^{2}{\alpha_{1}}) \right ] \\
   & + k_{2} r^{5}_{2} \left(\frac{\omega_{2}}{\overline{\omega}}\right)^{2} \left(1 + \frac{M_{1}}{M_{2}}\right) 
     \left [ \cos{\alpha_{2}}(\cos{\alpha_{2}} - \cos{\beta_{2}}\cos{i}) + \frac{1}{2}\sin^{2}{i} (1 - 5\cos^{2}{\alpha_{2}}) \right ] \bigg\} \\
\end{aligned}
\end{equation}
 where g(e) is equivalent to Equation \ref{eq:ge}, 
$\dot{\omega}_{gr}$ is the precession rate due to general relativity, 
$\overline{\omega}$ is the mean motion ($\overline{\omega} = \frac{2\pi}{P}$),
$\omega_{1}$ and $\omega_{2}$ are the spin angular velocities (angular frequencies) of each star, 
$k_{1}$ and $k_{2}$ are the apsidal motion constants of each star, 
$r_{1}$ and $r_{2}$ are the radii of each star, 
$M_{1}$ and $M_{2}$ are the masses of each star, 
$i$ is the inclination of the orbit, 
$e$ is the eccentricity of the orbit,
$\alpha_{1}$ and $\alpha_{2}$ are the angles between each star's axis of rotation and normal to the orbital plane, and 
$\beta_{1}$ and $\beta_{2}$ are the angles between each star's axis of rotation and the line of sight to the observer from the center of the binary system. 
For systems with $i \approx 90^{\circ}$ the contribution to the precession rate due to variations in the $\beta$ angles is negligible, so we focus on variations caused by changing $\alpha_{1}$ and $\alpha_{2}$.

Three cases were examined to see how important the effect of spin alignment is: 
the KOI-126 B and C (inner) orbit, DI Her, and BW Aqr. The last two binary star systems were included as controls, as this effect is known to be important in DI Her and not important in BW Aqr. 
System parameters for DI Her were taken from \citet{2009Natur.461..373A} and from \citet{1991AA...246..397C} for BW Aqr.

\begin{figure}
\centering
\includegraphics[width=0.5\linewidth]{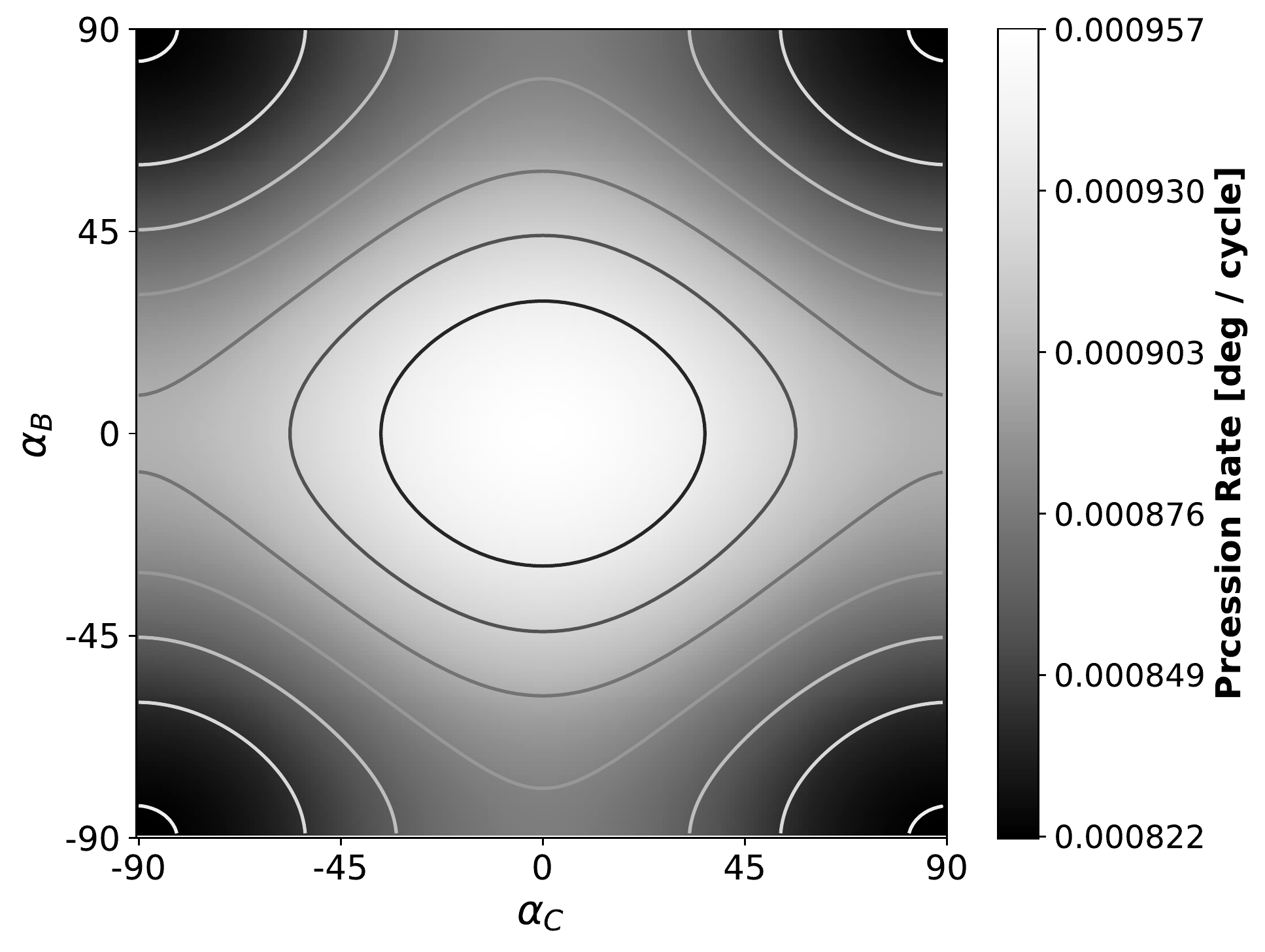}
\caption{Grayscale image showing the precession rate of the KOI-126 B and C binary 
if KOI-126 A did not exist using the stellar and orbital parameters from Table \ref{tab:final-parms} 
and the theoretical $k_{2}$ values \citep{2011ApJ...740L..25F}. 
The apsidal precession rate depends on the angle between 
the star's rotational axis with the normal of orbital plane ($\alpha$) \citep{ 1985SvAL...11..224S}. 
The maximum precession rate occurs when the stellar rotational axes of each star are 
aligned with the normal of orbital plane ($\alpha = 0^{\circ}$), 
while the minimum occurs when they are perpendicular to each other ($\alpha = \pm 90^{\circ}$).  
The contours were added to better see the landscape of the grids and are not represented by the color bar.}
\label{fig:shakura}
\end{figure}

Each star has its own value for the $\alpha$ angle, so a 2-d grid of $\dot{\omega}$ for each combination of the two possible $\alpha_{1}$ and $\alpha_{2}$ values was generated for each binary system. Equation \ref{eq:shakura} only considers the spin axis orientation and not the direction of the spin (i.e.\ prograde or retrograde), so values between -90$^{\circ}$ and 90$^{\circ}$ were used. 
Figure \ref{fig:shakura} shows the apsidal precession rate as a function of the $\alpha$ angles
for KOI-126 B and C. Notice that the fastest precession occurs when both stars have their spin aligned with the normal of the orbital plane, and the precession is slowest when both stars have their spin axes in the orbital plane. The symmetry in the figure is due to the fact that both stars have nearly equal masses and radii. For DI Her and BW Aqr, the figure looks qualitatively similar, except the scale of the precession rate differs and there is a greater dependence on the more massive star in the binary. 

As expected, we find the precession rate of DI Her can drop a very significant amount, from 0.00225 degrees/cycles to 0.00030 degrees/cycles (an 86\% decrease) due to the tilt of the stars' spin axes. 
The precession rate of DI Her of 0.00042 degrees/cycle corresponds to $\alpha_{1} = 72^{\circ}$ and $\alpha_{2} = -84^{\circ}$ \citep{2009Natur.461..373A} (note that the definition of $\alpha$ in Equation \ref{eq:shakura} is equivalent to the $\beta$ angle used by \citet{2009Natur.461..373A}).
For BW Aqr the axes of rotation play a much smaller role in the overall precession rate. With a maximum precession rate of 0.00089 deg/cycle when the $\alpha_{1,2} = 0^{\circ}$ and a minimum of 0.00084 degrees/cycle when $\alpha_{1,2} = \pm90^{\circ}$, the largest possible decrease in the precession rates is only 6\%. 

For KOI-126 B and C, using the theoretical $k_{2}$ values of 0.149 and 0.151 from \cite{2013ApJ...765...86F} gives an apsidal precession rate of 0.00098 deg/cycle (20.8 deg/century) when the stars' spin and orbit axes are aligned ($\alpha_{B,C} = 0^{\circ}$). This is the maximum binary precession rate possible given the B and C system parameters. When misaligned ($\alpha_{B,C} = \pm90^{\circ}$), the precession rate decreases to 0.00084 deg/cycles (17.8 deg/century), a 15\% drop. This is the minimum expected precession rate. Yet it is still significantly larger than the precession rate of 0.00046 deg/cycle (9.7 deg/century) when our observed value of $\overline{k_{2}} = 0.046$ is used. This is assuming aligned spin and orbit axes, otherwise the disagreement would be even larger. 
We thus conclude that 
misalignment of the spin axes cannot reduce the precession enough so as to make a $\overline{k_{2}}$ value of 0.15 appear to be a $\overline{k_{2}}$ value of 0.046. 

To summarize, in our modeling we have ignored the spin axis angles, which is equivalent to assuming they are in perfect alignment with the orbital axis, and hence the precession occurs at the maximum rate. If the stellar axes were misaligned, then the model would over-predict the rate of precession. Then, to match the observations, the models would reduce the $k_2$ values. However, from this investigation we find that any spin-orbit misalignment, if present, is insufficient to explain our observed lower-than-expected $\overline{k_{2}}$ value.

\subsubsection{The Effects of Non-sphericity, Spin Rates, Age and Metallicity}\label{sec:discussion-misc}
Due to tidal forces and stellar rotation, each star in KOI-126 is slightly non-spherical. This is accounted for in the equations of motion and the dynamics of the orbits. However, while calculating the model eclipse light curves the stars are assumed to be perfectly spherical. While there is no indication of a systematic trend in the light curve residuals, we checked the level of non-sphericity by using ELC in its ``numerical mode'' rather than its ``analytic mode''. In numerical mode, the stars' shapes follow their Roche equipotentials (see \cite{2005ApJ...632.1157W} and also \cite{1979ApJ...234.1054W}). Using this Roche geometry for the system parameters of the B+C inner binary, the polar and both equatorial radii of the stars (towards the other star and in the direction of motion) differ by only a few hundredths of a percent. This difference is negligible in the light curve construction and is therefore very unlikely to bias the solution towards a lower $\overline{k_{2}}$. 

The force equations also depend on the ratio of spin to orbital frequency, and we assume KOI-126 is in pseudo-synchronous rotation \citep{1981AA....99..126H}, as expected for a system with such a short orbital period. However, we have no independent observational evidence to support this assumption. If each stellar component were spinning slower than the expected pseudo-synchronous rate, the precession rate would be dampened and lower values for $\overline{k_{2}}$ would be inferred. By setting the stellar rotation frequency in Equation \ref{eq:k2cij} to zero we can find the apsidal precession rate from Equation \ref{eq:k2period} if neither star is spinning. Using the theoretical values for $k_{2}$ of 0.149 and 0.151 with
non-spinning stars gives an $\dot{\omega}$ rate of 0.00088 deg/cycle (=18.8 deg/century), compared to the pseudo-synchronously spinning stars with $\dot{\omega}$ of 0.00098 deg/cycle (=20.8 deg/century). This reduction is not enough to account for the inferred precession rate of 0.00046 deg/cycle (=9.7 deg/century) using the observed $\overline{k_{2}}$ value.

Since the apsidal motion constant depends on the internal mass distribution inside the star, as the star evolves the value for $k_{2}$ will also evolve. In principle, $k_2$ could be used as an age and metalicity diagnostic \citep{2013ApJ...765...86F}, and this is especially true for rapidly evolving high-mass stars (e.g.\ see \cite{2020A&A...642A.221R} and references therein). For very low-mass stars such as KOI-126 B and C, this evolution is so slow that it is generally not important once the star settles on the main sequence: \cite{2013ApJ...765...86F} show that $k_2 = 0.15$ at an age of $\sim$200 Myr. Even if KOI-126 B and C were only 1 Myr in age, the value of $k_2$ would be $\sim$0.13, instead of 0.15 \citep{2013ApJ...765...86F}. A similar argument could be made based on the metalicity of the stars but, again, the greatest difference expected occurs at an age of 1 Myr and leads to $k_{2} \approx$ 0.1325 at a low metallicity ([Fe/H] = -0.5) and $k_{2} \approx$ 0.1275 at a high metallicity ([Fe/H] = +0.2). At ages of 1 Gyr and greater the metalicity of the system has little bearing on the value of $k_{2}$ \citep{2013ApJ...765...86F}.

The conclusion of these subsections is that despite the low value for $\overline{k_{2}}$ that we infer, we can find no reason to think it may be in error. We do note that the effect of changing $\overline{k_{2}}$ from 0 to 0.6 (the upper limit found by \cite{2011Sci...331..562C}) creates a readily measurable change in the model light curve over the span of our observations. In contrast, a change from 0 to 0.15 produces a much more subtle effect on the model light curve. This is exacerbated 
by an inexact
 knowledge of the system: the masses, inclinations, radii, limb darkening, etc., are parameters that can vary slightly to compensate for an inaccurate $k_2$ value. But as simulations have shown, the input $k_2$ values are recoverable within the uncertainties. In the next section we examine what we suspect is the dominant source of error and appreciate the difficulty of measuring $k_2$ in a close triple-star system. 

In contrast, a change from 0 to 0.5 produces a much more subtle effect on the model light curve, this is exacerbated by an inexact knowledge of the system: the masses, inclinations, radii, limb darkening, etc., are parameters that can vary slightly to compensate for an inaccurate $k_{2}$ value.

\subsection{Third Body Effect vs.\ Tidal Apsidal Precession}\label{sec:3body}
Using the theoretical values of $k_{2,B}$ and $k_{2,C}$ as well as our 
observed stellar and orbital parameters we can calculate the precession 
rate due solely to the apsidal motion constants using Equation \ref{eq:k2} and compare
that to the observed total apsidal precession rate.
The theoretical tidal + rotational apsidal precession rate is 0.00076 deg/cycle (=16.1 deg/century) (N.B.\ GR precession contributes another 30\% to the precession, i.e.\ an additional 0.00022 deg/cycle (=4.6 deg/century)). With one cycle of the inner binary being 1.722 days, one classical apsidal period is $\sim$830,000 days (2270 years), and over the course of our observation we only sample
a tiny fraction (0.4\%) of this orbit.
However, the observations show that the inner orbit precesses at an astonishingly quick pace, with at one apsidal cycle every 1.741 years. This rapid precession is due to three-body gravitational dynamics, and this completely dominates over the classical and GR apsidal motion.
This rapid precession is both a blessing and a curse. The rapid precession allow us to put exquisitely tight constraints on the masses and radii and the orbital parameters, even though this is a single-lined spectroscopic system. The price we pay is that our hopes for measuring the internal mass distribution via the classical (tidal+rotational) apsidal motion to high precision is not possible. The very rapid precession that initially suggested we could measure $k_2$ to better than 1\% accuracy is in fact not caused by tides or rotation. Given the small fraction of the precession that is caused by the tides and rotation ($\sim$ 1329 time weaker then the 3-body effects), it is surprising that we are able to constrain $\overline{k_{2}}$ as well as we do.

\subsection{O--C of KOI-126 B and C}\label{sec:o-c}
In this final section we discuss a challenge presented by KOI-126 that normally is not a concern.
The classic method of measuring the apsidal motion constants is through the analysis of the  observed-minus-computed (O--C) residual curves of the eclipse times. This requires the precise determination of eclipse times {\em{independent of the binary system model.}} Normally this is not too difficult, as the mid-eclipse time is straightforward to measure and does not rely on precisely knowing the impact parameter or masses or any system parameter. This is not the case for KOI-126's complex eclipses (see Figure \ref{fig:kepler}). We do not observe stars B and C eclipsing each other, rather we observe these stars eclipse star A. The M stars sometimes eclipse the F star in a prograde sense and sometime retrograde. Consequently the depths of the eclipses change, the durations change, and the eclipses often overlap, creating complex shapes whose mid-eclipse times cannot be easily measured or interpreted. 

\begin{figure}[ht!]
\centering
\includegraphics[width=0.5\linewidth]{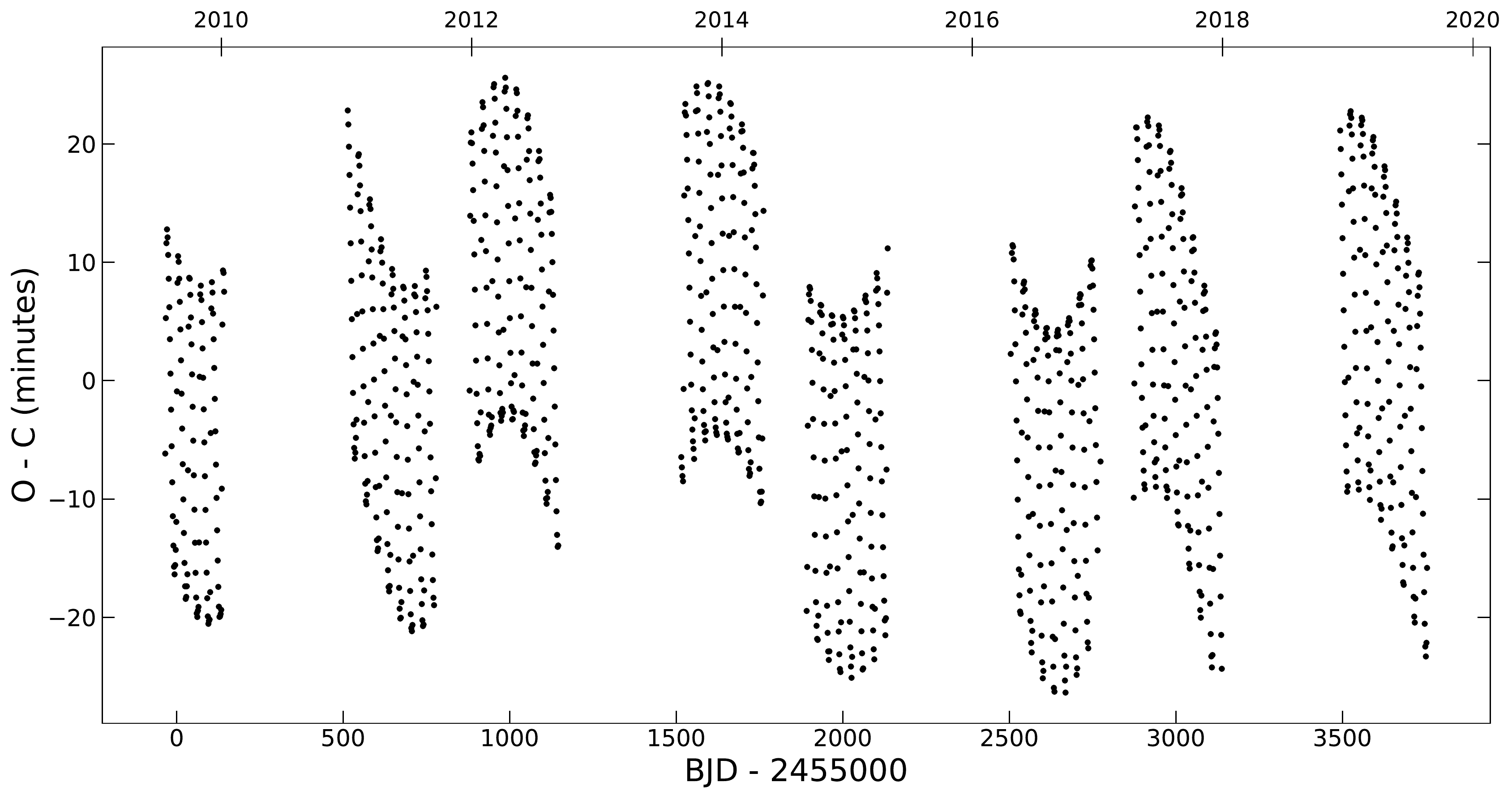}
\caption{Theoretical O-C curve for inner orbit eclipses. Eclipse times were taken from the dynamical
integration of the best fitting model, which was then fit linearly using least squares. 
The resulting ephemeris has a reference time of 2454999.989 BJD and a period of 1.750344 days, which was subtracted from the dynamical eclipse times.  Gaps occur when KOI-126 B and C are not eclipsing due to precession of the inner orbit inclination. Variations are primarily due to light travel time and third body dynamics, with small contributions from general relativistic and tidal effects.}
\label{fig:O-C}
\end{figure}

However, for comparison purposes, it is possible to compute what the expected O-C curve for KOI-126 B and C may look like if the eclipses were readily apparent (i.e.\ turn off the light from star A).
The primary eclipse (ie. star C eclipses B) times for the inner orbit were extracted from a dynamical integration of the system. A linear fit to the eclipse times resulted in a reference epoch of 2454999.989 BJD and a period of 1.750344 days. This fit was then subtracted from the eclipse times resulting in Figure \ref{fig:O-C}. We see a much more complex structure compared to the typical sinusoidal O-C diagram.  There are large gaps where the stars are no longer eclipsing because the orbit has precessed out of alignment with the line-of-sight to the Earth, and there are rapid up-down periodic variations due to the light travel time effect as the B+C binary orbits around the system barycenter. The remaining variations are due to a combination of changes in the inclination, argument of periastron, and nodal angle due to 3-body dynamics (and to a much smaller extent due to GR and tidal+rotational effects).

\section{Summary}\label{sec:conclusion}

We revisit the compact hierarchical triple-star system KOI-126, in which a pair of late M stars in a $\sim$1.7 d orbit revolve around an F star every $\sim$34 days. The F star, known as KOI-126 A, is eclipsed by stars B and C, but the duration of the eclipse is a significant fraction of the B+C orbital period. Consequently, the stars can exhibit significant orbital motion during the eclipse, and sometimes transit the star in a prograde or retrograde manner, leading to very complex and variable eclipse profiles.

In their discovery paper, \cite{2011Sci...331..562C} used \ik~observations of 8 eclipses
and two occultations
spanning 247 days. In this study we utilize the full 4-year \ik~data set that contains 42 eclipses, 
40 occultation windows,
and spans 1455 days. We also use ground-based eclipse observations that increase the baseline 5 years and now span a range 14 times larger than used in the \cite{2011Sci...331..562C} study. In addition, we add 13 more radial velocity measurements to the 16 in the discovery paper, which increases the RV baseline by 6 years. These increases in the observed time span are crucial, since the apsidal precession timescale for the B+C orbit is a remarkably brief $\sim$1.75 years and these additional data now sample several complete cycles.

Using photodynamical modeling to simultaneously fit the eclipse photometry and RV data, 
we arrive at stellar masses of 
M$_{A} = 1.2713   \pm 0.0047  ~M_{\odot}$,
M$_{B} = 0.23529 \pm 0.00062 ~M_{\odot}$, and
M$_{C} = 0.20739 \pm 0.00055 ~M_{\odot}$
and radii of 
R$_{A} = 1.9984 \pm 0.0027 ~R_{\odot}$,
R$_{B} = 0.25504 \pm 0.00076 ~R_{\odot}$, and
R$_{C} = 0.23196 \pm 0.00069 ~R_{\odot}$.
These are among the most precise stellar mass and radius determinations, especially for the 
M-stars. Only CM Dra has comparable precision \citep{2009ApJ...691.1400M}. The stars in the KOI-126 system can be well-matched with a stellar isochrone of age of $\sim$4.5 Gyrs, and like \cite{2011Sci...331..562C}, we find the M stars do not show the usual inflated-radius discrepancy common to M stars (e.g.\ \citet{2010AARv..18...67T}, \citet{2013ApJ...776...87S}).

Spurred on by the prediction in \cite{2011Sci...331..562C} that the apsidal motion constant $k_2$ for stars B and C could be measured to a precision of $\sim$1\% when the full \ik~data set is used, we have attempted to measure the $k_2$ values. Determining $k_2$ is important because it provides a constraint on the radial mass distribution inside the star ($k_2$ is akin to the Love number or the polytropic index). For M stars, there is no other known way to observationally constrain the internal mass distribution (unlike asteroseismology for high mass stars). Such information would be extremely valuable for testing stellar evolution models, which as noted above, often are not able to match the observed radii of M stars. Because the stars B and C are so similar in mass and radius, there is degeneracy in the solution for $k_2$ and we do not trust the individual estimates, but their sum (or mean) is constrained. Using our photodynamical model to fit the observed eclipses (not just the individual eclipse times), we estimate a mean $\overline{k_{2}} = 0.046^{+0.046}_{-0.028}$. 
This is 2-$\sigma$ lower than the theoretically expected value of 0.15 from \citet{2013ApJ...765...86F}. We explore several possible explanations for this mild discrepancy, including spin axis mis-alignment, and conclude that the most likely cause is that the classical apsidal precession due to tides and rotational oblateness is completely dwarfed by gravitational 3-body effects, i.e., the presence of KOI-126 A dominates the orbital precession of the B+C pair.

\acknowledgments

This material is based upon work supported by the National Science Foundation under Grant No.\ NSF AST-1617004.
We thank John Hood, Jr., for his generous support of exoplanet research at SDSU and his stimulating questions.
Some of the  calculations presented here were performed at the RRZ of the
Universit\"at Hamburg, at the H\"ochstleistungs Rechenzentrum Nord (HLRN).

\newpage

\section{Appendix}\label{sec:appendix}

Table \ref{tab:etimes1} and \ref{tab:etimes2} show future eclipse ingress, mid-eclipse, and egress times in BJD. In addition the absolute value of the predicted impact parameter and duration of each eclipse is given. The uncertainties are generated from posterior distributions of 121,540 posterior models taken from the final DEMCMC run described in Section \ref{sec:methods}.
Syzygy events may be of particular interest for photometric and spectroscopic observations so the start, end, and duration to the syzygy events are given in Table \ref{tab:syz}. These times were generated from dynamical integrations with a time step of 7.2 minutes, so this is a lower limit on the uncertainty of the timing accuracy.

\begin{deluxetable}{ccccccc}
\tabletypesize{\scriptsize}
\tablecaption{Predicted KOI-126 Eclipse Times (2021 - 2024)}
\tablewidth{0pt}
\label{tab:etimes1}
\tablehead{
\colhead{Star} &
\colhead{UTC Date} &
\colhead{Ingress} &
\colhead{Mid Eclipse} &
\colhead{Egress} &
\colhead{Duration}&
\colhead{Impact Parameter} \\
\colhead{ } &
\colhead{ } &
\colhead{BJD}  &
\colhead{BJD} &
\colhead{BJD} &
\colhead{Hours} &
\colhead{ } \\
}
\renewcommand{\arraystretch}{0.65}
\startdata
C & 01/12/2021 & 2459226.93017 $\pm$ 0.00138 & 2459227.25832 $\pm$ 0.00052 & 2459227.42667 $\pm$ 0.00034 & 11.916 $\pm$ 0.028 & 0.4152 $\pm$ 0.0010 \\
B & 01/12/2021 & 2459226.95903 $\pm$ 0.00033 & 2459227.03058 $\pm$ 0.00034 & 2459227.10481 $\pm$ 0.00036 & 3.499 $\pm$ 0.003 & 0.7592 $\pm$ 0.0005 \\
B & 02/15/2021 & 2459260.59222 $\pm$ 0.00041 & 2459260.75411 $\pm$ 0.00058 & 2459261.02057 $\pm$ 0.00116 & 10.28 $\pm$ 0.022 & 0.5466 $\pm$ 0.0010 \\
C & 02/15/2021 & 2459260.88752 $\pm$ 0.00035 & 2459260.96159 $\pm$ 0.00033 & 2459261.03269 $\pm$ 0.00033 & 3.484 $\pm$ 0.004 & 0.7328 $\pm$ 0.0007 \\
C & 03/20/2021 & 2459294.48354 $\pm$ 0.00034 & 2459294.56195 $\pm$ 0.00037 & 2459294.64893 $\pm$ 0.00043 & 3.969 $\pm$ 0.005 & 0.7286 $\pm$ 0.0007 \\
B & 03/21/2021 & 2459294.71225 $\pm$ 0.00052 & 2459294.84896 $\pm$ 0.00037 & 2459294.95893 $\pm$ 0.00033 & 5.920 $\pm$ 0.008 & 0.5881 $\pm$ 0.0010 \\
C & 04/23/2021 & 2459328.18774 $\pm$ 0.00074 & 2459328.56693 $\pm$ 0.00113 & 2459328.83584 $\pm$ 0.00043 & 15.554 $\pm$ 0.010 & 0.3815 $\pm$ 0.0012 \\
B & 04/23/2021 & 2459328.41082 $\pm$ 0.00035 & 2459328.48422 $\pm$ 0.00035 & 2459328.55847 $\pm$ 0.00037 & 3.544 $\pm$ 0.003 & 0.7376 $\pm$ 0.0006 \\
B & 05/27/2021 & 2459362.01442 $\pm$ 0.00040 & 2459362.14512 $\pm$ 0.00048 & 2459362.32719 $\pm$ 0.00077 & 7.506 $\pm$ 0.013 & 0.5232 $\pm$ 0.0009 \\
C & 05/27/2021 & 2459362.33755 $\pm$ 0.00038 & 2459362.41231 $\pm$ 0.00035 & 2459362.48231 $\pm$ 0.00035 & 3.474 $\pm$ 0.004 & 0.7671 $\pm$ 0.0007 \\
C & 06/30/2021 & 2459395.93056 $\pm$ 0.00036 & 2459396.00793 $\pm$ 0.00037 & 2459396.09081 $\pm$ 0.00042 & 3.846 $\pm$ 0.004 & 0.6969 $\pm$ 0.0007 \\
B & 06/30/2021 & 2459396.09321 $\pm$ 0.00086 & 2459396.26596 $\pm$ 0.00047 & 2459396.39066 $\pm$ 0.00037 & 7.139 $\pm$ 0.016 & 0.6407 $\pm$ 0.0011 \\
C & 08/03/2021 & 2459429.56619 $\pm$ 0.00055 & 2459429.79423 $\pm$ 0.00096 & 2459430.18758 $\pm$ 0.00079 & 14.913 $\pm$ 0.012 & 0.3943 $\pm$ 0.0011 \\
B & 08/03/2021 & 2459429.86650 $\pm$ 0.00037 & 2459429.93920 $\pm$ 0.00037 & 2459430.01081 $\pm$ 0.00038 & 3.463 $\pm$ 0.003 & 0.7543 $\pm$ 0.0006 \\
B & 09/05/2021 & 2459463.45518 $\pm$ 0.00041 & 2459463.56598 $\pm$ 0.00045 & 2459463.70235 $\pm$ 0.00057 & 5.932 $\pm$ 0.007 & 0.5454 $\pm$ 0.0009 \\
C & 09/06/2021 & 2459463.78674 $\pm$ 0.00043 & 2459463.85968 $\pm$ 0.00038 & 2459463.92611 $\pm$ 0.00038 & 3.345 $\pm$ 0.006 & 0.8244 $\pm$ 0.0006 \\
C & 10/09/2021 & 2459497.38400 $\pm$ 0.00037 & 2459497.45844 $\pm$ 0.00039 & 2459497.53589 $\pm$ 0.00042 & 3.645 $\pm$ 0.004 & 0.6995 $\pm$ 0.0007 \\
B & 10/09/2021 & 2459497.40080 $\pm$ 0.00168 & 2459497.64655 $\pm$ 0.00076 & 2459497.79829 $\pm$ 0.00046 & 9.540 $\pm$ 0.034 & 0.7188 $\pm$ 0.0012 \\
C & 11/12/2021 & 2459530.98581 $\pm$ 0.00050 & 2459531.13773 $\pm$ 0.00064 & 2459531.39987 $\pm$ 0.00127 & 9.937 $\pm$ 0.023 & 0.4836 $\pm$ 0.0011 \\
B & 11/12/2021 & 2459531.32198 $\pm$ 0.00039 & 2459531.39215 $\pm$ 0.00039 & 2459531.45971 $\pm$ 0.00041 & 3.306 $\pm$ 0.003 & 0.7944 $\pm$ 0.0006 \\
B & 12/16/2021 & 2459564.90534 $\pm$ 0.00041 & 2459565.00111 $\pm$ 0.00044 & 2459565.10947 $\pm$ 0.00050 & 4.899 $\pm$ 0.005 & 0.6052 $\pm$ 0.0009 \\
C & 12/16/2021 & 2459565.23346 $\pm$ 0.00049 & 2459565.30233 $\pm$ 0.00044 & 2459565.36333 $\pm$ 0.00044 & 3.117 $\pm$ 0.007 & 0.8875 $\pm$ 0.0005 \\
B & 01/19/2022 & 2459598.67651 $\pm$ 0.00124 & 2459598.92114 $\pm$ 0.00155 & 2459599.16496 $\pm$ 0.00071 & 11.723 $\pm$ 0.020 & 0.7665 $\pm$ 0.0013 \\
C & 01/19/2022 & 2459598.84289 $\pm$ 0.00039 & 2459598.91322 $\pm$ 0.00040 & 2459598.98440 $\pm$ 0.00043 & 3.396 $\pm$ 0.003 & 0.7275 $\pm$ 0.0007 \\
C & 02/21/2022 & 2459632.42605 $\pm$ 0.00049 & 2459632.53889 $\pm$ 0.00054 & 2459632.68862 $\pm$ 0.00073 & 6.302 $\pm$ 0.010 & 0.6007 $\pm$ 0.0011 \\
B & 02/22/2022 & 2459632.77139 $\pm$ 0.00040 & 2459632.84050 $\pm$ 0.00042 & 2459632.90564 $\pm$ 0.00045 & 3.222 $\pm$ 0.004 & 0.8263 $\pm$ 0.0006 \\
B & 03/27/2022 & 2459666.35800 $\pm$ 0.00043 & 2459666.44182 $\pm$ 0.00044 & 2459666.53170 $\pm$ 0.00049 & 4.169 $\pm$ 0.004 & 0.6753 $\pm$ 0.0008 \\
C & 03/28/2022 & 2459666.66594 $\pm$ 0.00055 & 2459666.73421 $\pm$ 0.00053 & 2459666.79271 $\pm$ 0.00059 & 3.043 $\pm$ 0.009 & 0.9290 $\pm$ 0.0005 \\
B & 04/30/2022 & 2459700.05274 $\pm$ 0.00087 & 2459700.20199 $\pm$ 0.00100 & 2459700.43978 $\pm$ 0.00132 & 9.289 $\pm$ 0.021 & 0.7999 $\pm$ 0.0011 \\
C & 04/30/2022 & 2459700.30108 $\pm$ 0.00040 & 2459700.36924 $\pm$ 0.00042 & 2459700.43644 $\pm$ 0.00045 & 3.249 $\pm$ 0.003 & 0.7524 $\pm$ 0.0007 \\
C & 06/03/2022 & 2459733.87824 $\pm$ 0.00049 & 2459733.96417 $\pm$ 0.00051 & 2459734.06460 $\pm$ 0.00058 & 4.473 $\pm$ 0.007 & 0.7158 $\pm$ 0.0009 \\
B & 06/03/2022 & 2459734.20716 $\pm$ 0.00044 & 2459734.28301 $\pm$ 0.00045 & 2459734.35228 $\pm$ 0.00049 & 3.483 $\pm$ 0.005 & 0.8255 $\pm$ 0.0007 \\
B & 07/07/2022 & 2459767.80999 $\pm$ 0.00044 & 2459767.88512 $\pm$ 0.00046 & 2459767.96300 $\pm$ 0.00049 & 3.672 $\pm$ 0.004 & 0.7304 $\pm$ 0.0007 \\
C & 07/07/2022 & 2459768.04641 $\pm$ 0.00077 & 2459768.14463 $\pm$ 0.00070 & 2459768.22008 $\pm$ 0.00082 & 4.168 $\pm$ 0.017 & 0.9252 $\pm$ 0.0008 \\
B & 08/09/2022 & 2459801.47406 $\pm$ 0.00075 & 2459801.56698 $\pm$ 0.00079 & 2459801.68884 $\pm$ 0.00102 & 5.155 $\pm$ 0.015 & 0.8715 $\pm$ 0.0008 \\
C & 08/10/2022 & 2459801.74962 $\pm$ 0.00043 & 2459801.82185 $\pm$ 0.00044 & 2459801.89121 $\pm$ 0.00047 & 3.398 $\pm$ 0.004 & 0.7432 $\pm$ 0.0007 \\
C & 09/12/2022 & 2459835.33419 $\pm$ 0.00050 & 2459835.40270 $\pm$ 0.00051 & 2459835.47742 $\pm$ 0.00055 & 3.437 $\pm$ 0.005 & 0.7950 $\pm$ 0.0008 \\
B & 09/13/2022 & 2459835.61842 $\pm$ 0.00054 & 2459835.71702 $\pm$ 0.00051 & 2459835.80144 $\pm$ 0.00053 & 4.392 $\pm$ 0.007 & 0.7800 $\pm$ 0.0008 \\
B & 10/16/2022 & 2459869.25954 $\pm$ 0.00047 & 2459869.33163 $\pm$ 0.00048 & 2459869.40442 $\pm$ 0.00050 & 3.477 $\pm$ 0.003 & 0.7499 $\pm$ 0.0006 \\
C & 10/16/2022 & 2459869.29983 $\pm$ 0.00148 & 2459869.51410 $\pm$ 0.00106 & 2459869.63724 $\pm$ 0.00114 & 8.098 $\pm$ 0.030 & 0.8922 $\pm$ 0.0012 \\
B & 11/19/2022 & 2459902.90798 $\pm$ 0.00067 & 2459902.97192 $\pm$ 0.00071 & 2459903.04582 $\pm$ 0.00082 & 3.308 $\pm$ 0.010 & 0.9149 $\pm$ 0.0005 \\
C & 11/19/2022 & 2459903.18422 $\pm$ 0.00049 & 2459903.26841 $\pm$ 0.00048 & 2459903.34654 $\pm$ 0.00048 & 3.896 $\pm$ 0.004 & 0.6935 $\pm$ 0.0008 \\
C & 12/23/2022 & 2459936.78732 $\pm$ 0.00052 & 2459936.84797 $\pm$ 0.00052 & 2459936.91172 $\pm$ 0.00055 & 2.986 $\pm$ 0.004 & 0.8271 $\pm$ 0.0007 \\
B & 12/23/2022 & 2459936.98445 $\pm$ 0.00084 & 2459937.13497 $\pm$ 0.00062 & 2459937.24857 $\pm$ 0.00058 & 6.339 $\pm$ 0.012 & 0.6906 $\pm$ 0.0010 \\
C & 01/26/2023 & 2459970.54154 $\pm$ 0.00093 & 2459970.75177 $\pm$ 0.00616 & 2459970.97705 $\pm$ 0.00291 & 10.452 $\pm$ 0.056 & 0.9259 $\pm$ 0.0017 \\
B & 01/26/2023 & 2459970.70471 $\pm$ 0.00051 & 2459970.78133 $\pm$ 0.00051 & 2459970.85658 $\pm$ 0.00051 & 3.645 $\pm$ 0.003 & 0.7268 $\pm$ 0.0007 \\
B & 02/28/2023 & 2460004.33927 $\pm$ 0.00061 & 2460004.39818 $\pm$ 0.00066 & 2460004.46360 $\pm$ 0.00078 & 2.984 $\pm$ 0.008 & 0.9035 $\pm$ 0.0005 \\
C & 03/01/2023 & 2460004.60479 $\pm$ 0.00060 & 2460004.70886 $\pm$ 0.00053 & 2460004.80021 $\pm$ 0.00051 & 4.690 $\pm$ 0.006 & 0.6205 $\pm$ 0.0008 \\
C & 04/03/2023 & 2460038.23643 $\pm$ 0.00055 & 2460038.29671 $\pm$ 0.00054 & 2460038.35868 $\pm$ 0.00056 & 2.934 $\pm$ 0.004 & 0.8213 $\pm$ 0.0006 \\
B & 04/03/2023 & 2460038.26305 $\pm$ 0.00147 & 2460038.52284 $\pm$ 0.00086 & 2460038.68408 $\pm$ 0.00067 & 10.105 $\pm$ 0.024 & 0.5802 $\pm$ 0.0012 \\
C & 05/07/2023 & 2460071.89540 $\pm$ 0.00070 & 2460071.98527 $\pm$ 0.00113 & 2460072.11341 $\pm$ 0.00239 & 5.232 $\pm$ 0.046 & 0.8944 $\pm$ 0.0010 \\
B & 05/07/2023 & 2460072.14407 $\pm$ 0.00059 & 2460072.23097 $\pm$ 0.00054 & 2460072.31341 $\pm$ 0.00053 & 4.064 $\pm$ 0.004 & 0.6804 $\pm$ 0.0007 \\
B & 06/10/2023 & 2460105.77199 $\pm$ 0.00060 & 2460105.83668 $\pm$ 0.00065 & 2460105.90698 $\pm$ 0.00074 & 3.240 $\pm$ 0.006 & 0.8525 $\pm$ 0.0006 \\
C & 06/10/2023 & 2460106.00383 $\pm$ 0.00086 & 2460106.14170 $\pm$ 0.00062 & 2460106.25066 $\pm$ 0.00054 & 5.924 $\pm$ 0.011 & 0.5470 $\pm$ 0.0009 \\
B & 07/14/2023 & 2460139.51184 $\pm$ 0.00123 & 2460139.85117 $\pm$ 0.00138 & 2460140.09482 $\pm$ 0.00087 & 13.992 $\pm$ 0.013 & 0.4992 $\pm$ 0.0013 \\
C & 07/14/2023 & 2460139.68480 $\pm$ 0.00058 & 2460139.74857 $\pm$ 0.00057 & 2460139.81279 $\pm$ 0.00057 & 3.072 $\pm$ 0.004 & 0.7957 $\pm$ 0.0006 \\
C & 08/16/2023 & 2460173.29510 $\pm$ 0.00065 & 2460173.38192 $\pm$ 0.00085 & 2460173.49009 $\pm$ 0.00130 & 4.680 $\pm$ 0.020 & 0.8146 $\pm$ 0.0011 \\
B & 08/17/2023 & 2460173.57629 $\pm$ 0.00069 & 2460173.67560 $\pm$ 0.00060 & 2460173.76603 $\pm$ 0.00056 & 4.554 $\pm$ 0.006 & 0.6474 $\pm$ 0.0009 \\
B & 09/19/2023 & 2460207.21022 $\pm$ 0.00061 & 2460207.28049 $\pm$ 0.00064 & 2460207.35486 $\pm$ 0.00071 & 3.472 $\pm$ 0.005 & 0.7977 $\pm$ 0.0006 \\
C & 09/19/2023 & 2460207.34161 $\pm$ 0.00179 & 2460207.55754 $\pm$ 0.00081 & 2460207.69346 $\pm$ 0.00059 & 8.444 $\pm$ 0.032 & 0.4941 $\pm$ 0.0011 \\
B & 10/23/2023 & 2460240.85905 $\pm$ 0.00092 & 2460241.09267 $\pm$ 0.00180 & 2460241.45039 $\pm$ 0.00152 & 14.192 $\pm$ 0.018 & 0.5109 $\pm$ 0.0016 \\
C & 10/23/2023 & 2460241.13516 $\pm$ 0.00062 & 2460241.20277 $\pm$ 0.00059 & 2460241.26913 $\pm$ 0.00059 & 3.215 $\pm$ 0.004 & 0.7772 $\pm$ 0.0007 \\
C & 11/26/2023 & 2460274.72036 $\pm$ 0.00063 & 2460274.80577 $\pm$ 0.00076 & 2460274.90565 $\pm$ 0.00099 & 4.447 $\pm$ 0.012 & 0.7390 $\pm$ 0.0011 \\
B & 11/26/2023 & 2460274.99864 $\pm$ 0.00087 & 2460275.11158 $\pm$ 0.00068 & 2460275.20916 $\pm$ 0.00060 & 5.053 $\pm$ 0.010 & 0.6529 $\pm$ 0.0010 \\
C & 12/30/2023 & 2460308.51934 $\pm$ 0.00255 & 2460308.92679 $\pm$ 0.00138 & 2460309.11779 $\pm$ 0.00071 & 14.363 $\pm$ 0.047 & 0.4721 $\pm$ 0.0015 \\
B & 12/30/2023 & 2460308.65267 $\pm$ 0.00064 & 2460308.72559 $\pm$ 0.00066 & 2460308.80071 $\pm$ 0.00071 & 3.553 $\pm$ 0.004 & 0.7668 $\pm$ 0.0007 \\
B & 02/01/2024 & 2460342.25926 $\pm$ 0.00082 & 2460342.41361 $\pm$ 0.00116 & 2460342.67390 $\pm$ 0.00240 & 9.951 $\pm$ 0.041 & 0.5255 $\pm$ 0.0013 \\
C & 02/02/2024 & 2460342.58708 $\pm$ 0.00068 & 2460342.65614 $\pm$ 0.00064 & 2460342.72206 $\pm$ 0.00062 & 3.240 $\pm$ 0.005 & 0.7932 $\pm$ 0.0007 \\
C & 03/06/2024 & 2460376.16279 $\pm$ 0.00064 & 2460376.24434 $\pm$ 0.00072 & 2460376.33480 $\pm$ 0.00087 & 4.128 $\pm$ 0.008 & 0.7035 $\pm$ 0.0010 \\
B & 03/06/2024 & 2460376.40186 $\pm$ 0.00125 & 2460376.53498 $\pm$ 0.00084 & 2460376.64061 $\pm$ 0.00067 & 5.730 $\pm$ 0.018 & 0.6983 $\pm$ 0.0011 \\
C & 04/09/2024 & 2460409.82571 $\pm$ 0.00138 & 2460410.15078 $\pm$ 0.00237 & 2460410.50347 $\pm$ 0.00102 & 16.266 $\pm$ 0.013 & 0.4443 $\pm$ 0.0018 \\
B & 04/09/2024 & 2460410.09968 $\pm$ 0.00066 & 2460410.17194 $\pm$ 0.00068 & 2460410.24447 $\pm$ 0.00073 & 3.475 $\pm$ 0.004 & 0.7707 $\pm$ 0.0007 \\
B & 05/13/2024 & 2460443.68423 $\pm$ 0.00079 & 2460443.80752 $\pm$ 0.00096 & 2460443.97085 $\pm$ 0.00141 & 6.879 $\pm$ 0.018 & 0.5521 $\pm$ 0.0011 \\
C & 05/13/2024 & 2460444.03905 $\pm$ 0.00075 & 2460444.10485 $\pm$ 0.00070 & 2460444.16622 $\pm$ 0.00069 & 3.052 $\pm$ 0.006 & 0.8467 $\pm$ 0.0007 \\
C & 06/16/2024 & 2460477.61444 $\pm$ 0.00066 & 2460477.69066 $\pm$ 0.00072 & 2460477.77144 $\pm$ 0.00083 & 3.768 $\pm$ 0.006 & 0.7148 $\pm$ 0.0010 \\
B & 06/16/2024 & 2460477.75208 $\pm$ 0.00253 & 2460477.93284 $\pm$ 0.00123 & 2460478.05401 $\pm$ 0.00082 & 7.246 $\pm$ 0.046 & 0.7702 $\pm$ 0.0012 \\
C & 07/19/2024 & 2460511.22164 $\pm$ 0.00107 & 2460511.40591 $\pm$ 0.00162 & 2460511.78590 $\pm$ 0.00232 & 13.542 $\pm$ 0.035 & 0.5057 $\pm$ 0.0013 \\
B & 07/20/2024 & 2460511.55201 $\pm$ 0.00069 & 2460511.62056 $\pm$ 0.00072 & 2460511.68758 $\pm$ 0.00077 & 3.254 $\pm$ 0.005 & 0.8060 $\pm$ 0.0008 \\
B & 08/22/2024 & 2460545.12503 $\pm$ 0.00078 & 2460545.22750 $\pm$ 0.00088 & 2460545.34839 $\pm$ 0.00110 & 5.361 $\pm$ 0.010 & 0.6173 $\pm$ 0.0011 \\
C & 08/22/2024 & 2460545.48909 $\pm$ 0.00081 & 2460545.54630 $\pm$ 0.00080 & 2460545.59890 $\pm$ 0.00083 & 2.635 $\pm$ 0.008 & 0.9109 $\pm$ 0.0006 \\
B & 09/25/2024 & 2460578.98133 $\pm$ 0.00380 & 2460579.25235 $\pm$ 0.00296 & 2460579.42915 $\pm$ 0.00122 & 10.748 $\pm$ 0.069 & 0.8390 $\pm$ 0.0016 \\
C & 09/25/2024 & 2460579.06809 $\pm$ 0.00068 & 2460579.13861 $\pm$ 0.00074 & 2460579.21082 $\pm$ 0.00082 & 3.426 $\pm$ 0.005 & 0.7511 $\pm$ 0.0010 \\
C & 10/29/2024 & 2460612.64915 $\pm$ 0.00096 & 2460612.77128 $\pm$ 0.00117 & 2460612.94892 $\pm$ 0.00188 & 7.194 $\pm$ 0.026 & 0.6271 $\pm$ 0.0012 \\
B & 10/29/2024 & 2460613.00520 $\pm$ 0.00071 & 2460613.06928 $\pm$ 0.00077 & 2460613.13041 $\pm$ 0.00084 & 3.005 $\pm$ 0.006 & 0.8503 $\pm$ 0.0008 \\
B & 12/02/2024 & 2460646.57599 $\pm$ 0.00078 & 2460646.66088 $\pm$ 0.00085 & 2460646.75404 $\pm$ 0.00098 & 4.273 $\pm$ 0.007 & 0.7043 $\pm$ 0.0010 \\
C & 12/02/2024 & 2460646.92877 $\pm$ 0.00081 & 2460646.97783 $\pm$ 0.00095 & 2460647.02227 $\pm$ 0.00113 & 2.244 $\pm$ 0.013 & 0.9537 $\pm$ 0.0007 \\
\enddata
\end{deluxetable}

\begin{deluxetable}{ccccccc}
\tabletypesize{\scriptsize}
\tablecaption{Predicted KOI-126 Eclipse Times (2025 - 2029)}
\tablewidth{0pt}
\label{tab:etimes2}
\tablehead{
\colhead{Star} &
\colhead{UTC Date} &
\colhead{Ingress} &
\colhead{Mid Eclipse} &
\colhead{Egress} &
\colhead{Duration}&
\colhead{Impact Parameter} \\
\colhead{ } &
\colhead{ } &
\colhead{BJD}  &
\colhead{BJD} &
\colhead{BJD} &
\colhead{Hours} &
\colhead{ } \\
}
\renewcommand{\arraystretch}{0.65}
\startdata
B & 01/04/2025 & 2460680.30359 $\pm$ 0.00223 & 2460680.46440 $\pm$ 0.00249 & 2460680.71214 $\pm$ 0.00231 & 9.805 $\pm$ 0.030 & 0.8659 $\pm$ 0.0017 \\
C & 01/05/2025 & 2460680.51937 $\pm$ 0.00070 & 2460680.58555 $\pm$ 0.00076 & 2460680.65157 $\pm$ 0.00084 & 3.173 $\pm$ 0.005 & 0.7848 $\pm$ 0.0009 \\
C & 02/07/2025 & 2460714.09098 $\pm$ 0.00093 & 2460714.17899 $\pm$ 0.00102 & 2460714.28375 $\pm$ 0.00124 & 4.627 $\pm$ 0.011 & 0.7482 $\pm$ 0.0012 \\
B & 02/07/2025 & 2460714.44596 $\pm$ 0.00077 & 2460714.51216 $\pm$ 0.00084 & 2460714.57365 $\pm$ 0.00093 & 3.064 $\pm$ 0.007 & 0.8669 $\pm$ 0.0009 \\
B & 03/13/2025 & 2460748.02855 $\pm$ 0.00080 & 2460748.10080 $\pm$ 0.00085 & 2460748.17634 $\pm$ 0.00093 & 3.547 $\pm$ 0.005 & 0.7742 $\pm$ 0.0008 \\
C & 03/13/2025 & 2460748.32931 $\pm$ 0.00094 & 2460748.39146 $\pm$ 0.00123 & 2460748.44401 $\pm$ 0.00160 & 2.753 $\pm$ 0.027 & 0.9586 $\pm$ 0.0011 \\
B & 04/16/2025 & 2460781.70506 $\pm$ 0.00169 & 2460781.79123 $\pm$ 0.00176 & 2460781.90707 $\pm$ 0.00219 & 4.848 $\pm$ 0.029 & 0.9212 $\pm$ 0.0010 \\
C & 04/16/2025 & 2460781.96504 $\pm$ 0.00076 & 2460782.03179 $\pm$ 0.00080 & 2460782.09668 $\pm$ 0.00086 & 3.159 $\pm$ 0.005 & 0.7924 $\pm$ 0.0009 \\
C & 05/20/2025 & 2460815.54026 $\pm$ 0.00093 & 2460815.60513 $\pm$ 0.00097 & 2460815.67619 $\pm$ 0.00106 & 3.262 $\pm$ 0.009 & 0.8405 $\pm$ 0.0010 \\
B & 05/20/2025 & 2460815.85808 $\pm$ 0.00096 & 2460815.94319 $\pm$ 0.00096 & 2460816.01838 $\pm$ 0.00102 & 3.847 $\pm$ 0.009 & 0.8313 $\pm$ 0.0011 \\
B & 06/22/2025 & 2460849.47414 $\pm$ 0.00086 & 2460849.54211 $\pm$ 0.00088 & 2460849.61099 $\pm$ 0.00092 & 3.284 $\pm$ 0.004 & 0.7998 $\pm$ 0.0007 \\
C & 06/23/2025 & 2460849.62797 $\pm$ 0.00159 & 2460849.76654 $\pm$ 0.00182 & 2460849.85834 $\pm$ 0.00230 & 5.529 $\pm$ 0.048 & 0.9352 $\pm$ 0.0018 \\
B & 07/26/2025 & 2460883.12884 $\pm$ 0.00141 & 2460883.17666 $\pm$ 0.00142 & 2460883.23047 $\pm$ 0.00157 & 2.439 $\pm$ 0.019 & 0.9606 $\pm$ 0.0006 \\
C & 07/26/2025 & 2460883.39992 $\pm$ 0.00088 & 2460883.47670 $\pm$ 0.00086 & 2460883.54865 $\pm$ 0.00087 & 3.569 $\pm$ 0.005 & 0.7565 $\pm$ 0.0009 \\
C & 08/29/2025 & 2460916.99049 $\pm$ 0.00096 & 2460917.04377 $\pm$ 0.00096 & 2460917.09985 $\pm$ 0.00099 & 2.625 $\pm$ 0.007 & 0.8798 $\pm$ 0.0007 \\
B & 08/29/2025 & 2460917.22608 $\pm$ 0.00145 & 2460917.35708 $\pm$ 0.00117 & 2460917.46020 $\pm$ 0.00111 & 5.619 $\pm$ 0.015 & 0.7481 $\pm$ 0.0013 \\
C & 10/02/2025 & 2460950.84471 $\pm$ 0.00277 & 2460951.07860 $\pm$ 0.00302 & 2460951.20602 $\pm$ 0.00568 & 8.673 $\pm$ 0.187 & 0.9564 $\pm$ 0.0035 \\
B & 10/02/2025 & 2460950.91031 $\pm$ 0.00096 & 2460950.98274 $\pm$ 0.00093 & 2460951.05427 $\pm$ 0.00094 & 3.455 $\pm$ 0.004 & 0.7807 $\pm$ 0.0007 \\
B & 11/05/2025 & 2460984.54738 $\pm$ 0.00121 & 2460984.59084 $\pm$ 0.00128 & 2460984.63781 $\pm$ 0.00143 & 2.171 $\pm$ 0.012 & 0.9542 $\pm$ 0.0005 \\
C & 11/05/2025 & 2460984.81773 $\pm$ 0.00112 & 2460984.91653 $\pm$ 0.00097 & 2460985.00359 $\pm$ 0.00091 & 4.460 $\pm$ 0.008 & 0.6802 $\pm$ 0.0009 \\
C & 12/08/2025 & 2461018.43659 $\pm$ 0.00100 & 2461018.49077 $\pm$ 0.00098 & 2461018.54647 $\pm$ 0.00098 & 2.637 $\pm$ 0.006 & 0.8668 $\pm$ 0.0007 \\
B & 12/08/2025 & 2461018.49000 $\pm$ 0.00295 & 2461018.74013 $\pm$ 0.00165 & 2461018.89097 $\pm$ 0.00129 & 9.623 $\pm$ 0.045 & 0.6432 $\pm$ 0.0014 \\
C & 01/11/2026 & 2461052.10810 $\pm$ 0.00128 & 2461052.17491 $\pm$ 0.00243 & 2461052.26494 $\pm$ 0.00513 & 3.764 $\pm$ 0.102 & 0.9567 $\pm$ 0.0015 \\
B & 01/11/2026 & 2461052.33991 $\pm$ 0.00110 & 2461052.42317 $\pm$ 0.00101 & 2461052.50264 $\pm$ 0.00096 & 3.905 $\pm$ 0.006 & 0.7336 $\pm$ 0.0008 \\
B & 02/14/2026 & 2461085.96441 $\pm$ 0.00114 & 2461086.01906 $\pm$ 0.00122 & 2461086.07765 $\pm$ 0.00135 & 2.718 $\pm$ 0.009 & 0.9068 $\pm$ 0.0006 \\
C & 02/14/2026 & 2461086.20874 $\pm$ 0.00165 & 2461086.34491 $\pm$ 0.00118 & 2461086.45284 $\pm$ 0.00099 & 5.858 $\pm$ 0.018 & 0.5949 $\pm$ 0.0010 \\
B & 03/20/2026 & 2461119.69218 $\pm$ 0.00227 & 2461120.05150 $\pm$ 0.00266 & 2461120.29587 $\pm$ 0.00175 & 14.488 $\pm$ 0.018 & 0.5686 $\pm$ 0.0023 \\
C & 03/20/2026 & 2461119.87936 $\pm$ 0.00107 & 2461119.93988 $\pm$ 0.00102 & 2461120.00067 $\pm$ 0.00100 & 2.911 $\pm$ 0.005 & 0.8324 $\pm$ 0.0007 \\
C & 04/22/2026 & 2461153.48629 $\pm$ 0.00118 & 2461153.55920 $\pm$ 0.00161 & 2461153.64758 $\pm$ 0.00242 & 3.871 $\pm$ 0.034 & 0.8825 $\pm$ 0.0014 \\
B & 04/23/2026 & 2461153.76332 $\pm$ 0.00133 & 2461153.86229 $\pm$ 0.00112 & 2461153.95193 $\pm$ 0.00100 & 4.526 $\pm$ 0.010 & 0.6851 $\pm$ 0.0010 \\
B & 05/26/2026 & 2461187.39164 $\pm$ 0.00112 & 2461187.45611 $\pm$ 0.00119 & 2461187.52409 $\pm$ 0.00130 & 3.179 $\pm$ 0.007 & 0.8454 $\pm$ 0.0007 \\
C & 05/27/2026 & 2461187.52579 $\pm$ 0.00373 & 2461187.75104 $\pm$ 0.00159 & 2461187.88856 $\pm$ 0.00112 & 8.706 $\pm$ 0.066 & 0.5335 $\pm$ 0.0013 \\
B & 06/29/2026 & 2461221.02499 $\pm$ 0.00167 & 2461221.25163 $\pm$ 0.00356 & 2461221.63113 $\pm$ 0.00337 & 14.547 $\pm$ 0.044 & 0.5931 $\pm$ 0.0029 \\
C & 06/29/2026 & 2461221.32034 $\pm$ 0.00116 & 2461221.38623 $\pm$ 0.00109 & 2461221.45074 $\pm$ 0.00104 & 3.130 $\pm$ 0.006 & 0.8122 $\pm$ 0.0008 \\
C & 08/02/2026 & 2461254.90127 $\pm$ 0.00114 & 2461254.97660 $\pm$ 0.00139 & 2461255.06220 $\pm$ 0.00180 & 3.862 $\pm$ 0.019 & 0.8139 $\pm$ 0.0013 \\
B & 08/02/2026 & 2461255.17340 $\pm$ 0.00177 & 2461255.29377 $\pm$ 0.00131 & 2461255.39481 $\pm$ 0.00109 & 5.314 $\pm$ 0.019 & 0.6680 $\pm$ 0.0012 \\
C & 09/05/2026 & 2461288.62121 $\pm$ 0.00368 & 2461289.09230 $\pm$ 0.00318 & 2461289.30264 $\pm$ 0.00139 & 16.354 $\pm$ 0.057 & 0.4910 $\pm$ 0.0020 \\
B & 09/05/2026 & 2461288.82977 $\pm$ 0.00114 & 2461288.89875 $\pm$ 0.00119 & 2461288.96945 $\pm$ 0.00127 & 3.352 $\pm$ 0.006 & 0.8081 $\pm$ 0.0008 \\
B & 10/08/2026 & 2461322.41921 $\pm$ 0.00144 & 2461322.56508 $\pm$ 0.00214 & 2461322.80834 $\pm$ 0.00464 & 9.339 $\pm$ 0.079 & 0.5980 $\pm$ 0.0020 \\
C & 10/09/2026 & 2461322.76119 $\pm$ 0.00129 & 2461322.82904 $\pm$ 0.00118 & 2461322.89361 $\pm$ 0.00112 & 3.178 $\pm$ 0.008 & 0.8252 $\pm$ 0.0009 \\
C & 11/11/2026 & 2461356.33270 $\pm$ 0.00113 & 2461356.40709 $\pm$ 0.00131 & 2461356.48784 $\pm$ 0.00158 & 3.723 $\pm$ 0.013 & 0.7766 $\pm$ 0.0013 \\
B & 11/12/2026 & 2461356.55280 $\pm$ 0.00283 & 2461356.70642 $\pm$ 0.00170 & 2461356.82120 $\pm$ 0.00126 & 6.441 $\pm$ 0.041 & 0.7026 $\pm$ 0.0016 \\
C & 12/15/2026 & 2461389.94909 $\pm$ 0.00210 & 2461390.23681 $\pm$ 0.00436 & 2461390.66930 $\pm$ 0.00222 & 17.284 $\pm$ 0.010 & 0.4472 $\pm$ 0.0013 \\
B & 12/15/2026 & 2461390.27465 $\pm$ 0.00120 & 2461390.34278 $\pm$ 0.00124 & 2461390.41048 $\pm$ 0.00131 & 3.260 $\pm$ 0.006 & 0.8183 $\pm$ 0.0009 \\
B & 01/18/2027 & 2461423.84325 $\pm$ 0.00135 & 2461423.95727 $\pm$ 0.00171 & 2461424.10385 $\pm$ 0.00252 & 6.255 $\pm$ 0.030 & 0.6274 $\pm$ 0.0017 \\
C & 01/18/2027 & 2461424.20416 $\pm$ 0.00147 & 2461424.26861 $\pm$ 0.00132 & 2461424.32816 $\pm$ 0.00125 & 2.976 $\pm$ 0.011 & 0.8751 $\pm$ 0.0009 \\
C & 02/21/2027 & 2461457.77660 $\pm$ 0.00112 & 2461457.84629 $\pm$ 0.00128 & 2461457.91913 $\pm$ 0.00150 & 3.421 $\pm$ 0.011 & 0.7837 $\pm$ 0.0014 \\
B & 02/21/2027 & 2461457.81863 $\pm$ 0.00779 & 2461458.07630 $\pm$ 0.00285 & 2461458.21848 $\pm$ 0.00162 & 9.596 $\pm$ 0.153 & 0.7719 $\pm$ 0.0020 \\
C & 03/26/2027 & 2461491.35075 $\pm$ 0.00172 & 2461491.51077 $\pm$ 0.00256 & 2461491.84484 $\pm$ 0.00642 & 11.858 $\pm$ 0.116 & 0.5474 $\pm$ 0.0015 \\
B & 03/27/2027 & 2461491.72230 $\pm$ 0.00124 & 2461491.78423 $\pm$ 0.00133 & 2461491.84427 $\pm$ 0.00144 & 2.927 $\pm$ 0.008 & 0.8658 $\pm$ 0.0010 \\
B & 04/29/2027 & 2461525.28139 $\pm$ 0.00133 & 2461525.37529 $\pm$ 0.00155 & 2461525.48201 $\pm$ 0.00193 & 4.815 $\pm$ 0.016 & 0.6928 $\pm$ 0.0015 \\
C & 04/30/2027 & 2461525.65309 $\pm$ 0.00169 & 2461525.70147 $\pm$ 0.00159 & 2461525.74574 $\pm$ 0.00161 & 2.224 $\pm$ 0.018 & 0.9507 $\pm$ 0.0009 \\
B & 06/02/2027 & 2461558.99745 $\pm$ 0.00400 & 2461559.23848 $\pm$ 0.00512 & 2461559.55677 $\pm$ 0.00259 & 13.423 $\pm$ 0.044 & 0.7846 $\pm$ 0.0038 \\
C & 06/02/2027 & 2461559.22989 $\pm$ 0.00113 & 2461559.29181 $\pm$ 0.00129 & 2461559.35437 $\pm$ 0.00148 & 2.988 $\pm$ 0.011 & 0.8270 $\pm$ 0.0014 \\
C & 07/06/2027 & 2461592.78202 $\pm$ 0.00156 & 2461592.89146 $\pm$ 0.00192 & 2461593.03781 $\pm$ 0.00284 & 6.139 $\pm$ 0.034 & 0.6727 $\pm$ 0.0015 \\
B & 07/06/2027 & 2461593.16927 $\pm$ 0.00124 & 2461593.22107 $\pm$ 0.00144 & 2461593.27048 $\pm$ 0.00166 & 2.429 $\pm$ 0.013 & 0.9201 $\pm$ 0.0011 \\
B & 08/09/2027 & 2461626.72537 $\pm$ 0.00133 & 2461626.80282 $\pm$ 0.00149 & 2461626.88520 $\pm$ 0.00173 & 3.836 $\pm$ 0.012 & 0.7727 $\pm$ 0.0013 \\
B & 09/11/2027 & 2461660.36938 $\pm$ 0.00269 & 2461660.50522 $\pm$ 0.00317 & 2461660.73085 $\pm$ 0.00475 & 8.675 $\pm$ 0.062 & 0.8224 $\pm$ 0.0020 \\
C & 09/12/2027 & 2461660.68309 $\pm$ 0.00115 & 2461660.73801 $\pm$ 0.00133 & 2461660.79204 $\pm$ 0.00153 & 2.615 $\pm$ 0.011 & 0.8711 $\pm$ 0.0014 \\
C & 10/15/2027 & 2461694.22800 $\pm$ 0.00150 & 2461694.30536 $\pm$ 0.00168 & 2461694.39344 $\pm$ 0.00200 & 3.971 $\pm$ 0.016 & 0.7938 $\pm$ 0.0013 \\
B & 10/16/2027 & 2461694.60592 $\pm$ 0.00122 & 2461694.65144 $\pm$ 0.00162 & 2461694.69413 $\pm$ 0.00202 & 2.117 $\pm$ 0.023 & 0.9509 $\pm$ 0.0013 \\
B & 11/18/2027 & 2461728.17061 $\pm$ 0.00135 & 2461728.23498 $\pm$ 0.00148 & 2461728.30081 $\pm$ 0.00166 & 3.125 $\pm$ 0.010 & 0.8379 $\pm$ 0.0011 \\
B & 12/22/2027 & 2461761.79229 $\pm$ 0.00236 & 2461761.86178 $\pm$ 0.00243 & 2461761.94504 $\pm$ 0.00278 & 3.666 $\pm$ 0.030 & 0.9230 $\pm$ 0.0014 \\
C & 12/22/2027 & 2461762.12156 $\pm$ 0.00129 & 2461762.17832 $\pm$ 0.00143 & 2461762.23278 $\pm$ 0.00160 & 2.669 $\pm$ 0.010 & 0.8773 $\pm$ 0.0013 \\
C & 01/25/2028 & 2461795.67631 $\pm$ 0.00151 & 2461795.73304 $\pm$ 0.00160 & 2461795.79311 $\pm$ 0.00173 & 2.803 $\pm$ 0.010 & 0.8749 $\pm$ 0.0010 \\
B & 01/25/2028 & 2461796.00481 $\pm$ 0.00158 & 2461796.06866 $\pm$ 0.00194 & 2461796.12502 $\pm$ 0.00234 & 2.885 $\pm$ 0.027 & 0.9357 $\pm$ 0.0017 \\
B & 02/28/2028 & 2461829.61000 $\pm$ 0.00146 & 2461829.67047 $\pm$ 0.00152 & 2461829.73059 $\pm$ 0.00162 & 2.894 $\pm$ 0.007 & 0.8589 $\pm$ 0.0008 \\
B & 04/01/2028 & 2461863.23385 $\pm$ 0.00237 & 2461863.26071 $\pm$ 0.00211 & 2461863.28892 $\pm$ 0.00210 & 1.322 $\pm$ 0.036 & 0.9844 $\pm$ 0.0009 \\
C & 04/02/2028 & 2461863.53575 $\pm$ 0.00163 & 2461863.60936 $\pm$ 0.00161 & 2461863.67695 $\pm$ 0.00164 & 3.389 $\pm$ 0.009 & 0.8307 $\pm$ 0.0012 \\
C & 05/05/2028 & 2461897.11735 $\pm$ 0.00161 & 2461897.16622 $\pm$ 0.00161 & 2461897.21632 $\pm$ 0.00163 & 2.375 $\pm$ 0.008 & 0.9018 $\pm$ 0.0008 \\
B & 05/05/2028 & 2461897.32139 $\pm$ 0.00291 & 2461897.45681 $\pm$ 0.00261 & 2461897.55594 $\pm$ 0.00267 & 5.629 $\pm$ 0.030 & 0.8703 $\pm$ 0.0023 \\
B & 06/08/2028 & 2461931.03525 $\pm$ 0.00171 & 2461931.10627 $\pm$ 0.00164 & 2461931.17448 $\pm$ 0.00161 & 3.341 $\pm$ 0.007 & 0.8225 $\pm$ 0.0008 \\
B & 07/12/2028 & 2461964.65601 $\pm$ 0.00212 & 2461964.68182 $\pm$ 0.00197 & 2461964.70850 $\pm$ 0.00190 & 1.260 $\pm$ 0.021 & 0.9814 $\pm$ 0.0007 \\
C & 07/12/2028 & 2461964.91846 $\pm$ 0.00236 & 2461965.02899 $\pm$ 0.00191 & 2461965.12063 $\pm$ 0.00174 & 4.852 $\pm$ 0.019 & 0.7407 $\pm$ 0.0012 \\
B & 08/14/2028 & 2461998.48852 $\pm$ 0.00382 & 2461998.78722 $\pm$ 0.00374 & 2461998.96024 $\pm$ 0.00357 & 11.32 $\pm$ 0.041 & 0.8000 $\pm$ 0.0039 \\
C & 08/15/2028 & 2461998.54957 $\pm$ 0.00175 & 2461998.60295 $\pm$ 0.00165 & 2461998.65656 $\pm$ 0.00159 & 2.568 $\pm$ 0.009 & 0.8810 $\pm$ 0.0009 \\
B & 09/17/2028 & 2462032.44317 $\pm$ 0.00215 & 2462032.53671 $\pm$ 0.00185 & 2462032.62174 $\pm$ 0.00169 & 4.286 $\pm$ 0.014 & 0.7481 $\pm$ 0.0009 \\
B & 10/21/2028 & 2462066.06875 $\pm$ 0.00198 & 2462066.11370 $\pm$ 0.00193 & 2462066.16016 $\pm$ 0.00192 & 2.194 $\pm$ 0.010 & 0.9331 $\pm$ 0.0007 \\
C & 10/21/2028 & 2462066.21086 $\pm$ 0.00562 & 2462066.42611 $\pm$ 0.00263 & 2462066.55762 $\pm$ 0.00196 & 8.322 $\pm$ 0.091 & 0.6271 $\pm$ 0.0012 \\
B & 11/24/2028 & 2462099.73029 $\pm$ 0.00266 & 2462099.92979 $\pm$ 0.01195 & 2462100.26464 $\pm$ 0.00901 & 12.825 $\pm$ 0.158 & 0.8598 $\pm$ 0.0052 \\
C & 11/24/2028 & 2462099.97845 $\pm$ 0.00195 & 2462100.04299 $\pm$ 0.00175 & 2462100.10592 $\pm$ 0.00161 & 3.059 $\pm$ 0.011 & 0.8340 $\pm$ 0.0011 \\
C & 12/28/2028 & 2462133.58480 $\pm$ 0.00191 & 2462133.61500 $\pm$ 0.00225 & 2462133.64677 $\pm$ 0.00275 & 1.488 $\pm$ 0.031 & 0.9747 $\pm$ 0.0010 \\
B & 12/28/2028 & 2462133.82730 $\pm$ 0.00303 & 2462133.95507 $\pm$ 0.00222 & 2462134.06069 $\pm$ 0.00184 & 5.601 $\pm$ 0.031 & 0.6750 $\pm$ 0.0013 \\
C & 01/30/2029 & 2462167.27229 $\pm$ 0.00492 & 2462167.75099 $\pm$ 0.00502 & 2462167.97464 $\pm$ 0.00251 & 16.857 $\pm$ 0.061 & 0.5180 $\pm$ 0.0012 \\
B & 01/30/2029 & 2462167.48926 $\pm$ 0.00202 & 2462167.54778 $\pm$ 0.00196 & 2462167.60712 $\pm$ 0.00195 & 2.829 $\pm$ 0.007 & 0.8794 $\pm$ 0.0007 \\
B & 03/05/2029 & 2462201.08582 $\pm$ 0.00227 & 2462201.19048 $\pm$ 0.00345 & 2462201.33673 $\pm$ 0.00646 & 6.022 $\pm$ 0.104 & 0.8253 $\pm$ 0.0026 \\
C & 03/05/2029 & 2462201.40496 $\pm$ 0.00228 & 2462201.48220 $\pm$ 0.00193 & 2462201.55414 $\pm$ 0.00169 & 3.580 $\pm$ 0.017 & 0.7975 $\pm$ 0.0013 \\
C & 04/08/2029 & 2462234.99530 $\pm$ 0.00185 & 2462235.04603 $\pm$ 0.00209 & 2462235.09943 $\pm$ 0.00241 & 2.499 $\pm$ 0.017 & 0.9077 $\pm$ 0.0011 \\
B & 04/08/2029 & 2462235.14396 $\pm$ 0.00631 & 2462235.34991 $\pm$ 0.00300 & 2462235.48444 $\pm$ 0.00210 & 8.171 $\pm$ 0.104 & 0.6300 $\pm$ 0.0019 \\
C & 05/12/2029 & 2462268.58178 $\pm$ 0.00292 & 2462268.88515 $\pm$ 0.01174 & 2462269.33660 $\pm$ 0.00432 & 18.116 $\pm$ 0.036 & 0.5644 $\pm$ 0.0064 \\
B & 05/12/2029 & 2462268.91518 $\pm$ 0.00214 & 2462268.98064 $\pm$ 0.00205 & 2462269.04518 $\pm$ 0.00202 & 3.120 $\pm$ 0.008 & 0.8535 $\pm$ 0.0008 \\
\enddata
\end{deluxetable}

\begin{deluxetable}{cccc}
\tablecaption{Approximate Syzygy Times}
\tablewidth{0pt}
\label{tab:syz}
\tablehead{
\colhead{UTC Date} &
\colhead{Start} &
\colhead{End} &
\colhead{Duration} \\
\colhead{ } &
\colhead{BJD}  &
\colhead{BJD} &
\colhead{Minutes} 
}
\startdata
02/09/2020 & 2458889.12 & 2458889.14 & 28.8 \\
07/27/2020 & 2459058.005 & 2459058.055 & 72.0 \\
01/19/2022 & 2459598.855 & 2459598.905 & 72.0 \\
04/30/2022 & 2459700.385 & 2459700.425 & 57.6 \\
10/16/2022 & 2459869.3 & 2459869.315 & 21.6 \\
01/26/2023 & 2459970.805 & 2459970.855 & 72.0 \\
07/20/2024 & 2460511.67 & 2460511.7 & 43.2 \\
09/25/2024 & 2460579.05 & 2460579.105 & 79.2 \\
01/05/2025 & 2460680.57 & 2460680.61 & 57.6 \\
10/02/2025 & 2460951.0 & 2460951.045 & 64.8 \\
03/20/2026 & 2461119.92 & 2461119.94 & 28.8 \\
06/02/2027 & 2461559.24 & 2461559.29 & 72.0 \\
11/18/2027 & 2461728.175 & 2461728.18 & 7.2 \\
\enddata
\end{deluxetable}

\newpage
\bibliography{bib}{}
\bibliographystyle{aasjournal}

\end{document}